\documentclass[preprint2]{aastex}

\slugcomment{manuscript for AJ, version 3.0, 131130 = submit AJ}

\shorttitle{Radio-optical reference frame link}
\shortauthors{Zacharias \& Zacharias}

\begin{document}


\title{Radio$-$optical reference frame link using the US Naval
       Observatory astrograph and deep CCD imaging}



\author{N. Zacharias$^{1,3}$,
        M.I. Zacharias$^{2,3}$}

\email{nz@usno.navy.mil}

\affil{$^1$U.S.~Naval Observatory, 3450 Mass.Ave.~NW Washington, DC 20392; \\
       $^2$U.S.~Naval Observatory (contractor); \\
       $^3$Visiting astronomer, Cerro Tololo Inter-American Observatory and
          Kitt Peak National Observatory, National Optical Astronomy 
          Observatories, which are operated by the Association of Universities
          for Research in Astronomy, under contract with the National
          Science Foundation.}


\begin{abstract}
Between 1997 and 2004 several observing runs were conducted mainly
with the CTIO 0.9 m to image ICRF counterparts (mostly QSOs) in order
to determine accurate optical positions.
Contemporary to these deep CCD images the same fields were observed
with the US Naval Observatory (USNO) astrograph in the same bandpass.  
They provide accurate
positions on the Hipparcos/Tycho-2 system for stars in the 10 to 16 
magnitude range used as reference stars for the deep CCD imaging data.
Here we present final optical position results of 413 sources based on
reference stars obtained by dedicated astrograph observations
which were reduced following 2 different procedures.
These optical positions are compared to radio VLBI positions.
The current optical system is not perfectly aligned to the ICRF radio 
system with rigid body rotation angles of 3 to 5 mas (= 3 $\sigma$ level) 
found between them for all 3 axes.
Furthermore, statistically, the optical$-$radio position differences are 
found to exceed the total, combined, known errors in the observations.
Systematic errors in the optical reference star positions as well as 
physical offsets between the centers of optical and radio emissions are 
both identified as likely causes.  A detrimental, astrophysical, random
noise (DARN) component is postulated to be on about the 10 mas level.
If confirmed by future observations, this could severely limit the Gaia 
to ICRF reference frame alignment accuracy to an error of about 0.5 mas 
per coordinate axis with the current number of sources envisioned to 
provide the link. 
A list of 36 ICRF sources without the detection of an optical counterpart 
to a limiting magnitude of about R=22 is provided as well.
\end{abstract}


\section{Introduction}

The Hipparcos Catalogue \citep{hipcat} is currently the primary
optical reference frame.
It is linked to the defining radio International Celestial Reference
Frame (ICRF) by various methods with a dozen radio stars observed
with Very Long Baseline Interferometry (VLBI) having the greatest
weight \citep{k97}.
An improved version of the ICRF, the ICRF2 \citep{icrf2} now contains 
over 3400 compact, extragalactic sources on the same coordinate system,
which is assumed to be inertial.

The Hipparcos Celestial Reference frame (HCRF) is entirely based on space
mission observations and contains positions, proper motions and parallaxes
of over 100,000 bright stars on the milli-arcsecond (mas) level
(mas/yr for individual proper motions).  Zonal and magnitude dependent
errors in the Hipparcos Catalogue are estimated to be even smaller.
The HCRF is estimated to be accurate to 0.6 mas (1$\sigma$) in its 
alignment to the ICRF at the epoch of 1991.25 with an estimated 0.25 mas/yr 
(1$\sigma$) residual system rotation error per rigid body angle as compared 
to the ICRF \citep{k97}.

The Tycho-2 catalog is the first step of the densification of the optical 
reference frame to contain the 2.5 million brightest stars to visual magnitude
about 11 \citep{tycho2}.  Random position errors of Tycho-2 stars are about
7 mas per coordinate at V = 9 and epoch 1991, increasing for the
fainter stars.  The random error of Tycho-2 star proper motions is typically
about 2 mas/yr.  The Tycho-2 data are based on Hipparcos star tracker 
observations (at mean epoch of 1991.25) combined with over 140 ground-based 
star catalogs to provide sufficient leverage in epoch difference for precise 
proper motions.  The largest weight of the early epoch data in Tycho-2 comes
from the Astrographic Catalogue (AC) \citep{ac2000}, with positions of 
photographic plates at an epoch around 1900.  The AC plates cover an 
area of just over 2$^{\circ}$ by 2$^{\circ}$ and Hipparcos stars were 
used for their astrometric reductions.
Although great care has been taken in these reductions, remaining systematic
errors on the AC plates are estimated to be on the 100 mas level, maybe more
(Urban, private com.).  This results in possible systematic errors of
Tycho-2 proper motions on the 1 to 2 mas/yr level.

The goal of this paper is to examine the link between the defining
radio reference frame (ICRF) and the optical reference frame (HCRF)
by observing the optical counterparts of ICRF sources on the HCRF
using the Tycho-2 as reference star catalog in a 2-step process to
bridge the large magnitude difference.  While the deep CCD imaging
captures the optical counterparts, the USNO astrograph is used to
provide intermediate bright (12 to 16 mag) reference stars.

Besides possible systematic errors in the optical reference frame
(i.e.~Tycho-2) the assumption of coinciding centroids of the radio 
and optical centers of emissions of the extragalactic
sources is questioned.  Observed optical$-$radio position offsets 
are found to be larger than expected from the combination of all
known, random errors.  Clues for both cases, reference frame issues,
and physical offset between radio and optical centers of emission
are found.  If the later turns out to be a wide-spread issue on the 
several mas level this would have serious consequences for the
alignment accuracy of the future optical, celestial reference frame 
from the Gaia mission \citep{gaia} to the ICRF.
Our data are not precise enough to make that case; however, some
individual sources are found with relatively large radio-optical position
offsets likely on the 30 mas level, a few with even much larger offsets,
indicating a physical non-coincidence of the radio-optical centroids.

A possible physical offset between the radio and optical centroid
of ICRF sources was first discussed in connection with a correlation
to the radio structure index \citep{dasilva}.  Results from a similar,
extensive investigation of 300 optical counterparts were recently 
published by the Rio de Janeiro group \citep{riosurvey}. 
Direct differences between ICRF2 radio positions and optical counterpart
positions found in the Sloan Digital Sky Survey (SDSS) DR9 were studied 
for 1297 common sources in another recent paper \citep{orosz}.
Positions of 22 ICRF sources observed with the SOAR 4.2m and ESO 2.2m 
telescopes again show a correlation of optical$-$radio position offsets 
with radio structure index \citep{camargo}. 

Our research presented here is the largest and most accurate
such sample published so far related to the original ICRF sources
and can be considered as a follow-up to an earlier work based on
photographic plates for secondary reference stars \citep{rorf99}.
Preliminary results of a subset of these data in collaboration
with another group was published earlier \citep{a2003}.
The high precision of our current investigation could be achieved 
by using dedicated astrometric
observations along with contemporaneous astrograph observations, 
thus eliminating the degrading effect of proper motion errors
of intermediate bright link stars.
Partial, preliminary results of our research were presented at
several meetings, \citep{adela}, \citep{iau2003}, \citep{lowel2005}, 
\citep{iaus248}, \citep{iau2009} and \citep{iau2012}.  
It was not until recently that the final (4th) USNO CCD Astrograph
Catalog (UCAC4) was published \citep{ucac4} and its reduction
process could be adopted for the critical astrogaph data to finally
conclude with this paper a research program that lasted 
more than a decade.

\section{Observations}

Optical positions of ICRF counterparts were derived from a 2-step
observing process using the USNO astrograph and the CTIO 0.9 m telescope,
respectively.  For each 0.9 m run the same fields were 
observed with the astrograph within about a month.  
Thus hard-to-come-by accurate proper motions of anonymous medium bright
stars common to both sets of observations are not needed excluding a
major source of error for the derived optical positions of the ICRF
counterparts. 
The large number of high quality secondary reference stars provided
by these astrograph observations also allowed calibration of the 0.9 m
data for field angle distortions on the mas level.

For most observing runs both telescopes were located at the same 
site (Cerro Tololo) and the same bandpass (579 to 643 nm) was
used with custom filters.  This approach minimizes astrometric errors
from differential color refraction (DCR) of the anonymous reference stars 
common to both data sets.  The relatively narrow bandpass also minimizes
DCR between the ICRF target source and the set of reference stars. 

For a total of 413 sources, mainly in the southern hemisphere,
optical positions could be derived.
These include AGN, BL-Lac and QSO sources.  
No photometric observations were performed here.
The distribution of redshift 
as a function of approximate optical magnitude is presented in Fig.~1 
with data taken from the LQAC-2 catalog \citep{lqac2}.

\subsection{CTIO 0.9 m}

Table 1 gives a summary of the deep imaging observing runs used for this 
investigation.  All except one made use of the CTIO 0.9 m telescope with 
its standard 2k by 2k CCD and a scale of 0.4 arcsec/pixel.
For a single observing run
we used the 2.1m KPNO telescope with its 2k by 2k CCD and 0.3 arcsec/pixel
image scale. 
The last deep observing run ``J" (CTIO 0.9 m) was performed a few years
after most of the other runs to acquire data on ICRF targets which were
missed before or could use more observing.  
Typically about 4 deep exposures were taken per source and
observing run with several arcsec offsets applied between exposures to 
sample the field of view with different parts of the detector from one 
exposure to the other.
The target source was always close to the center of the CCD (within about
100 pixels).  Typical exposure times were between 120 and 600 sec, 
depending on seeing and target brightness.  No photometric
standards were observed, and observations sometimes were performed 
in non-photometric conditions.

In addition to the target fields a number of astrometric calibration 
fields were also observed in each run.  Overlapping frames with large
separations of about 1/3 of the field-of-view (FOV) were taken in a 2 by 2
or 3 by 3 pattern with typically 200 sec exposures.  All these fields
are at low galactic latitude to provide a high star density, but are
not too crowded in order to avoid large numbers of blended images.
These fields significantly enhance the number of residuals available
later in the astrometric reductions to produce field distortion maps.
They also often contain a detected ICRF counterpart and allow 
comparison of results between different observing runs.

\subsection{USNO astrograph}

The same telescope which produced the USNO CCD Astrgraph Catalog (UCAC)
was also used to image a square degree area around each ICRF target at
about the same epoch as the deep CCD imaging runs.  
Some observing statistics for this investigation are given in Table 2.
Properties of the astrograph and its camera are given in Table 3.
The imaging optics is the 5-element ``red lens" with a
design FOV of $9^{\circ}$ diameter, thus the entire 4k by 4k CCD area
is practically ``on the optical axis" with very uniform image quality.

Typically sets of 4 long (150 sec) and 4 short (30 sec) exposures 
were taken with the astrograph on one side of the pier, with an
offset of about 1/4 of the FOV in a 2 by 2 pattern centered on the target.
A similar set of exposures was taken with the telescope flipped to the
other side of the pier.  This approach compensates for possible constant,
magnitude dependent systematic errors of derived star positions.
For some fields fewer numbers of exposures were taken, however, always
aiming at a balance between East and West of pier observations.

Important to note is that although the same instrument was used, none of 
the astrograph observations used for this ICRF investigation are part of 
the published UCAC star catalogs.  The special observing performed for
this investigation has a much larger number of exposures per field than
the general UCAC survey data, the additional benefit of symmetric 
East and West of pier observations, and a somewhat increased 
exposure time for better signal-to-noise performance of faint stars.

\section{Data Processing}

\subsection{Astrograph data processing}

The pixel-to-pixel sensitivity variations of the 4k CCD are small and
could be well corrected with mean flats.  No separate bias corrections
were applied beyond what has been corrected by applying the dark
frame calibrations.
Image centers were obtained from fits of the calibrated pixel data
with a 5-parameter, 2-dimensional image profile model.  
For the UCAC2 a Gauss function profile model was used followed by
correction of $x,y$ data for the pixel-phase error \citep{ucac2}. 
For UCAC4 a better matching modified Lorentz profile was utilized
for the image profile model function, using fixed shape parameters
determined from pilot investigations but using only 5 free
parameters for individual star fits in the final reduction \citep{ucac3px}. 

The astrograph data were initially processed following the procedures
of the UCAC2 reduction pipeline \citep{ucac2}, hereafter called 
UCAC2-type.
Later the raw data were re-processed with methods adopted for UCAC4 
\citep{ucac4}, drawing upon the UCAC3 pipeline (\cite{ucac3a}) for 
most of the code, hereafter called UCAC4-type.
Note that in neither case actual UCAC2 or UCAC4 catalog data were
used for these reference stars, rather dedicated, extra astrograph 
observations were processed with respective pipelines.

Due to the poor charge transfer efficiency (CTE) of the 4k CCD used at
the astrograph, the camera was operated at a relatively warm temperature
of about $-18^{\circ}$ C to mitigate the problem.  The resulting high
dark current required frequently constructing mean dark frames for the
standard exposure times involved.  Nevertheless a significant coma-like
systematic error exists in the astrograph data for the $x$-coordinate.
Elaborate corrections were applied to reduce the initial systematic
errors on the 100 mas level to about 10 mas.  Detailed procedures to 
correct for CTE effects were different between the UCAC2-type and 
UCAC4-type reductions (see above cited papers).  

The models for these corrections
were developed by utilizing 2MASS \citep{2mass} reference stars for
differential astrometry, spanning the entire magnitude range available
with the astrograph data.  No proper motions of 2MASS stars were available.
However, the mean epoch difference between the 2MASS catalog and the 
astrograph data is typically about a year thus global galactic kinematic
effects are irrelevant here over scales of the size of an astrograph exposure.  
We were looking only for position differences
as a function of magnitude and $x,y$ coordinates on a $1^{\circ}$ scale.
The large number of residuals (typically thousands for a parameter space bin)
allowed the large noise (order 80 mas, 1$\sigma$) of individual star position
differences to be reduced to insignificant levels as compared to the
systematic error calibration goal of about 10 mas.
Utilizing East and West of pier observations for this investigation further 
reduces the remaining effects of the poor CTE in the astrograph data.

After applying these systematic error corrections to the $x,y$ centers
of fitted stars an astrometric solution was performed using the Tycho-2 
\citep{tycho2} as reference star catalog.  However, stars fainter than
about magnitude 11 were excluded, as these were found to be of lower 
astrometric quality than the rest of the catalog.
A linear plate model was used in these reductions after handling 
atmospheric refraction and aberration rigorously with our custom code.
In an iterative process a field distortion map was constructed from the
residuals and binned by $x,y$ coordinates.  A smoothed version of this
map was applied to the data before the final astrometric solution
with the same ``plate" model.
A secondary reference star catalog was then generated from the weighted
mean positions of stars from all astrograph observations of these ICRF
fields, covering the about 10 to 16 mag range in our bandpass.

\subsection{Deep CCD data reduction}

Raw data processing of the deep exposures was performed with IRAF.
Mean bias and dome flat frames were applied for calibration.
Each exposure was checked for image quality like 
astigmatism related to focus and mean image elongation related to
guiding effects.  The target source was identified using finding
charts and the source checked for blended images, cosmic ray hits
or other problems.  Notes were made and flags set to exclude individual
exposures of questionable quality.

For the astrometric reductions of the deep CCD exposures the secondary
reference star catalog from the astrograph observations was used.
Again, atmospheric refraction and aberration as function of $x,y$
and telescope pointing were handled rigorously.
A weighted least-squares reduction was performed, considering the
formal $x,y$ fit errors of individual stars, an error contribution
from the turbulence of the atmosphere (scaled by the inverse 
square-root of exposure time), and the formal errors of individual 
reference star position coordinates.

Various plate models were explored and a significant 3rd order optical
distortion found, as expected for this type of optics.  The $x,y$ data,
for each telescope and observing run separately, was then pre-corrected
for the mean optical distortion term and astrometric reductions repeated
with a linear model.  Residuals were then binned as function of $x,y$
coordinates in the focal plane and a smoothed map generated from all
data of an individual observing run.  
An example of such a field distortion pattern is shown in Fig.~2
presenting data from the CTIO 0.9 m telescope observing run ``q"
based on UCAC4-type reference stars.  Over 40,000 residuals from
279 exposures were used for this vector plot.
These corrections were applied
to the $x,y$ data of individual exposures and the astrometric reductions
repeated with the same model and distortion pre-corrections.
Due to degraded optical quality far from the optical axis, reference 
stars in the corners of the CCD were excluded from the final astrometric 
reductions.

\subsection{Total random error of optical positions of targets}

The total optical position error of our ICRF counterpart targets
per coordinate, and per individual deep CCD expoxure, $\sigma_{tot_{i}}$, 
from all known, random error contributions is

\[  \sigma_{tot_{i}} = \sqrt { 
               \sigma_{xyfit}^{2} + \sigma_{as}^{2}
              + (100 mas)^{2} \frac{100 sec}{t_{exp} [sec]} }
\]
Included here are the $x,y$ position fit error of the target source
on individual deep exposures, $\sigma_{xyfit}$, the contribution from 
the error propagation of the astrometric solution, 
$\sigma_{as}$, see below, and the error from the turbulence in the 
atmosphere for the target source.
The last term scales with exposure time, $t_{exp}$, depends slightly
on the angular distance of the target to the reference stars, and is 
correlated with the ``astrometric seeing".
An estimate is adopted here based on previous observational data
with the 0.9 m CTIO telescope \citep{sigatm}.
This term represents the random offset of the observed center of 
the target source w.r.t.~its true position on the grid of 
reference stars, caused by the turbulence in the atmosphere.
The total, random error of a mean target position from optical data, 
$\sigma_{tot}$, as given in the results tables below then is the 
weighted mean of the above over all contributing exposures.

Turbulence in the atmosphere of course also contributes to the
astrometric solution error by affecting the reference stars.
This and the random catalog position errors of the reference stars 
and the random $x,y$ fit errors of the reference stars are automatically 
included in the formula for the error contribution from the astrometric 
solution, $\sigma_{as}$ (spelled out here for the $\xi$ coordinate),

\[ \sigma_{as}^{2} = \sum_{i=1}^{k} \left( 
          \frac{\partial \xi}{\partial p_{i}} \right)^{2}
            \sigma_{p_{i}}^{2}
\]

The formula for the $\eta$ coordinate is similar, with
$\xi, \eta$ being the standard coordinates in the tangential plane.
Here $\sigma_{p_{i}}$ are the errors of the $i=1,k$ ``plate constants"
of the adopted astrometric model between the standard coordinates,
and the observed $x,y$ image center coordinates of those reference
stars.  All error sources mentioned above contribute to the astrometic
solution error.  The partial derivatives by plate constants depend on
the adopted model and are evaluated at the position of the target
extragalactic source on a given deep CCD exposure.
In other words, $\sigma_{as}$ is the error contribution from the
error propagation of the astrometric solution process in the
tangential plane, i.e.~the error associated with how well the
coordinate system represented by the reference stars in the field
of view is established at the point of the target source on a given 
deep exposure.

\section{Results}

Weighted mean optical positions (per observing run) of ICRF counterparts 
were derived by the procedures described above, excluding deep exposures 
of questionable quality.  For most sources UCAC2-type and UCAC4-type
reductions were performed (see section 3.1).  
The optical reference frame used is the
Tycho-2 catalog, excluding its faintest stars.
Thus the optical positions presented here should be on the HCRS.
These positions are compared to the radio positions using the 
ICRF2 \citep{icrf2} data.
The optical$-$radio position differences and associated errors
are analyzed further as described below.

\subsection{Optical position tables}

The primary results of this investigation are presented in Tables 4 and 5,
for the UCAC2 and UCAC4-type reductions, respectively.  The full tables
are available from the CDS in Strasbourg \citep{cdscat}, while a sample 
is given here.  Both tables are in the same format.
The tables contain 666 and 682 lines (position results), respectively
for a total of 413 unique sources.  Many sources were observed in more than
one observing run.  Not all sources were reduced with both types of 
secondary reference stars.

Columns 1 and 2 list the optical positions, RA and Dec [hms format], 
respectively for individual sources and observing runs as derived from 
this investigation. 
Those positions are on the Tycho-2 coordinate system believed to be on
the Hipparcos Celestial Reference Frame, which in turn is constructed
to be on the ICRF.
Columns 3 and 4 give the same position in decimal degree format.
Column 5 is the radial distance [mas] between the optical and radio position
based on columns 6 and 7, which
give the position differences optical minus radio,
$\Delta \alpha \cos \delta$, $\Delta \delta$, respectively [mas].
Data from the published ICRF2 catalog \cite{icrf2} were used for the
radio positions.

Columns 8 and 9 give the total optical position error per coordinate,
as discussed in section 3.3.
Particularly the $x,y$-fit errors can vary significantly
due to differences in signal-to-noise (S/N) ratios of observed sources
as well as the number and quality of available reference stars.
However, due to the circular symmetric nature of the image profile fit
function the $x$ coordinate and $y$ coordinate errors are equal.
Similarly, error propagation values from $\sigma_{as}$ are almost the 
same for both coordinates, leading to identical values in columns 8 and 9 
when rounded to mas.

Column 10 gives the ratio of columns 6 and 8, i.e.~the optical$-$radio
position difference for RA in unit of ``sigmas".  Column 11 gives the same
for the Dec component.
Column 12 gives the error of the radio position [mas] (per coordinate; 
the larger of the RA or Dec component is shown) from the ICRF2.
In almost all cases this error is negligible as compared to the optical
position errors.  A value of 99 indicates an unknown error.

Column 13 gives the number of deep exposures used for the mean, optical
position of this source and run.  Column 14 provides the average number
of secondary reference stars used for this field in the deep CCD reductions,
and column 15 lists the number of reference stars excluded (above the 
adopted 3$\sigma$ threshold of astrometric solution error). 
Column 16 gives the mean astrometric 
plate solution error [mas] from the individual deep exposures.
Column 17 gives the approximate S/N ratio of the optical counterpart
image on a single deep exposure.

Column 18 and 19 give the optical brightness and redshift, respectively,
taken from the LQAC-2 \citep{lqac2} catalog.  If available the R magnitude
is used, otherwise, in order of preference, I, V, or g.  
Zero in this column indicates that
no magnitude was available from the above options.
Column 20 gives the object type 1, 2 or 3 for AGN, BL-Lac, or QSO,
respectively, according to the 13th edition of \citep{veron13}. 

Column 21 lists the X-band radio structure index. 
A value of 1 indicates an (almost) pointlike source, while 5 means
a lot of structure is seen.  A value of 9 indicates an unknown radio
structure index.
A combination of 3 data sets are used, with the Bordeaux database
\citep{Bsix} as first pick, followed by a recently published table
\citep{bourda} and the ICRF-2 catalog data.

Column 22 is either 2 or 4 to indicate the UCAC2 or UCAC4-type solution.
This allows the user to merge and sort both tables if desired, while
retaining that information for each row.
Column 23 gives the B1950 coordinate-based name of the source taken
from the ICRF2, which is commonly used in the VLBI community.
To obtain the J2000-based, IAU-style name of a source, just use
columns 1 and 2 to the desired level of digits.

Column 24 contains the 1-letter name of the deep CCD observing run
(see Table 1), while column 25 is either ``r" for a regular radio-optical
reference frame source in our program, or ``c" indicating also an optical
calibration field (see above).
The flag in column 26 indicates the quality of the optical source image,
with possible values explained in Table 6.
The lable ``n" is used for 2 sources not being in the ICRF.
Radio positions were taken as follows for 0447$-$439 \citep{p1998} and
2316$-$423 (Roopesh, priv.com.).
Finally, column 27 contains remarks, e.g.~if the optical image
looks like a resolved galaxy or is a double, in which case the flux
ratio (flr =  target source / component) and separation (arcsec) are given.

\subsection{Empty fields}

Table 7 lists the ICRF sources where an optical counterpart
within an arcsec of the radio position could not be found
to a limiting magnitude of about R = 22 (in most cases, sometime
20 to 21 with Moon and/or poor seeing).
In some of those cases, particularly in crowded fields at low
galactic latitudes, a nearby point source was found, but accurate 
astrometry rules out the possibility of a positive match with
the ICRF source.  In most cases the position offset from the
optical object (likely foreground star) to the ICRF radio position
is given in the comment column of Table 7.
Some of these fields had more than 1 observing run attempt to find
an optical counterpart.  The area around ICRF source 0826$-$373 was targeted
as one of the field distortion calibration fields imaged during several
observing runs.

\subsection{Global system orientation}

It is possible that the optical coordinate system as represented
by Tycho-2 stars in our investigation at around epoch 2000 is no
longer perfectly aligned with the ICRF radio coordinate system.
Allowing for 3 Eulerian, rigid body, rotation angles (as transformation
between the 2 static coordinate systems) in a least-squares 
solution of our optical minus radio position differences we find marginal 
significant results.
No attempt has been made to also solve for a system rotation (change of
orientation with time) due to the short time baseline of our observing 
epochs.

For comparison some solutions were obtained with an additional fit
parameter, a constant offset in declination.  Most of our sources
are in the southern hemisphere and such an offset is conceivable,
perhaps introduced by remaining systematic errors as a function of
magnitude.  The difference in brightness between our primary
reference stars (Tycho-2) and the ICRF optical counterparts is
almost 10 magnitudes and a very small remaining magnitude equation
could produce such a global offset in declination.  If present, a similar
offset in RA is already covered by the 3rd orientation angle around
the $z$ coordinate axis.

Table 8 summarizes the results of various solutions, giving the
Eulerian angles w1 to w3, declination offset (if solved for),
and the errors of those parameters.
The first column gives the fit error followed by the number of degrees
of freedom (number of observation equations used minus number of
parameters to solve for). The last column distinguishes between
different solutions, with u and w indicating unweighted and weighted
solutions, respectively.  The number 2 or 4 indicates UCAC2-type or
UCAC4-type data.  The ``3sig" indicates that observation equations
with residuals above 3$\sigma$ are excluded and limits of 150 mas
and 30 mas are put on optical$-$radio offsets and radio position 
errors, respectively.  
Similarly the less restrictive ``4sig" excludes observation equations
only if above 4$\sigma$ and adopts 200 mas and 90 mas for the
selection criteria mentioned above.

Depending on selection criteria for these solutions one can get 
quite different results.  However, there seems to be a consistent
trend with about +5, $-$4, and +5 mas for the angles 1 to 3, 
and $-4$ mas for the declination offset, with formal errors
of about 1.5 mas each.
These values are slightly larger than expected from the alignment
tolerances of the HCRS to ICRF systems alone \citep{k97}
which give a spin error of 0.25 mas/yr (1$\sigma$) leading to an expected
offset of just over 2 mas over the approximately 9 year mean epoch 
difference between Hipparcos (1991) and our data (2000). 

For the following statistics we do not correct our optical$-$radio
position differences for this possible global orientation misalignment,
but rather use the data as given in our optical$-$radio catalog 
(samples shown in Table 4 and 5).

\section{Analysis of optical minus radio position differences}

\subsection{Precision of position differences}

Fig.~3 shows the distribution of our formal, optical position
errors of the RA coordinate for the UCAC2-type reductions.
These errors include the $x,y$ position fit error of the sources,
atmospheric turbulence and error propagation from reference
star errors and ``plate" solutions of both steps in the
reductions.  These errors are for the mean optical position
of a target source over all exposures of a given observing run.
Results for Dec and for the UCAC4-type reductions are almost identical.
The high precision of our observations is apparent with most sources
falling in the 10 to 15 mas bin.

In the following we exclude source observations by 3 criteria: 
optical minus radio position 
differences exceeding 100 mas (in either coordinate), fsigma (columns 7,8
of tables 4 and 5) exceeding 3.5, and low S/N sources ($\le$ 5).
With this we get a
root-mean-square (RMS) optical$-$radio position offset of 24.2 and 24.3 mas
for the RA and Dec component, respectively for the UCAC2-type solution.
In these statistics 501 out of 666 source observations of table 4 are
included (306 individual sources).  
The corresponding numbers for the UCAC4-type solution are
27.1 and 26.0 mas for 506 out of 682 source observations and
317 individual sources.
This result makes our investigation the most precise of its kind
published so far.

Fig.~4 shows a histogram of the UCAC2-type optical$-$radio position
differences of all of our data, while Fig.~5 shows the same for the 
UCAC4-type data.
Individual sources in this set have significantly different S/N
and thus significantly different image fit precisions and total errors.
Thus individual observations come from different statistical populations
and Figs.~4 and 5 can't be expected to be normal distributed.

\subsection{Underestimation of formal errors}

Figs.~6 and 7 show histograms of the same data as Figs.~4 and 5 but 
scaled for individual errors, i.e.~the distribution of (optical$-$radio)
position differences divided by the total, known (1$\sigma$) error for 
that observation.  If the error estimates were correct and the
position differences were normal-distributed, then the histogram
data would be consistent with a Gaussian function of standard deviation 1.0.
The curve plotted over the histograms in Figs.~6 and 7 are the
best fit Gauss functions, however the standard deviations
are 1.25 and 1.29 for the RA and Dec component of the UCAC2-type
data respectively, and 1.40 and 1.33 for the UCAC4-type data.
Here outliers were excluded, using 545 data points out of 666 and
541 out of 682 for the UCAC2 and UCAC4-type data, respectively.
Including these outliers the best fit distributions would be even
wider.

Thus these distributions are about 30\% wider than expected 
from known error sources.  Furthermore all distributions are 
``heavy tailed" even when compared to the already broader
best fit normal distribution.  A large number of sources are above
3$\sigma$, consistent with what is seen in other investigations
of this type.
As already indicated from the global system rotation analysis (see 
above) Figs.~6 and 7 also show that the means of these distributions are 
slightly offset from zero.

Table 9 lists the 88 observations of 63 sources with (optical$-$radio) 
position differences over 4.0 $\sigma$ (of total, formal error as listed
in Table 4) in either coordinate from our UCAC2-type data.  
This corresponds roughly to 3$\sigma$ outliers assuming the best fit
Gaussian distribution as shown in Fig.~6 and 7.
The format is the same as that of Table 4.
These include several ``problem" sources, as can be seen from the
comment column.  A few might even be ``no match" cases.
Some other entries in Table 9 are sources with moderate (optical$-$radio)
position offsets (50 to 100 mas) where very small, total, formal optical
position errors result in exceeding the 4$\sigma$ threshold.
The corresponding list from the UCAC4-type data (not shown here)
is very similar and can be obtained by selecting from the published
full data catalog \citep{cdscat} (sample shown in Table 5).

There are 2 possible causes which could explain the larger than expected
optical$-$radio position differences: systematic errors in the
optical reference star data, and astrophysical non-coincidence
of the optical and radio centers of emission.
As shown below there are strong indications for the presence
of both explanations.  Systematic errors in the radio reference
frame are much smaller than the 20 to 200 mas level needed to
explain the position discrepancies.

\subsection{Vector plots}

The key results of our main tables 4 and 5 are visualized in 
Fig.~8 and 9 showing the location of our sources on the sky together 
with vectors indicating the optical$-$radio position differences.
Fig.~8 shows the RA range of 0 to 12 hours, while Fig.~9 shows the
12 to 24 hours area (each with some overlap).
The scale of the vectors is 10 mas per degree.  The red and blue vectors
show the results of the UCAC2 and UCAC4-type reductions, respectively.
Excluded from these plots are sources with formal errors or position
differences exceeding 200 mas per coordinate.
The few sources with excessive position differences are discussed below.
In many local areas the vectors seem to be aligned, indicating systematic
errors related to the reference stars.

\subsection{Systematic errors in the optical data}

The primary optical reference frame used in this investigation is the
Tycho-2 \citep{tycho2} star catalog.  The mean positions of stars in
the Tycho-2 catalog are based on Hipparcos space mission observations
at central epoch 1991.25.  However, the Tycho-2 proper motions were 
derived in combination with many early epoch ground-based catalogs,
of which the Astrographic Catalogue (AC) project from the 1900s is of
particular importance because of the large epoch difference, its
high precision and coverage of the 9 to 11 mag range, i.e. the
stars used as primary reference stars in our investigation.

The AC data consists of some 20,000 photographic plates of about
2.1 by 2.1 degree size, taken in a 2-fold overlap pattern
(center in corner),
with plate centers on each full degree of declination \citep{eichhorn}.
These AC measures were reduced with great care using Hipparcos
reference stars \citep{ac2000}.
However, removing systematic errors, particularly those depending
on magnitude and products of magnitude and $x,y$ coordinate (coma terms)
was limited to about 100 to 200 mas (Urban, private comm.)
The corresponding error floor of the Tycho-2 proper motions thus is
in the order of 1 to 2 mas/yr, possibly varying over scales of 1 degree
(distance from center to edge on an AC plate).
This translates to remaining, typical, systematic errors from the Tycho-2
catalog of 10 to 20 mas for reference stars at our mean 2000 
epoch of optical counterpart observations (9 years after the Tycho-2
central epoch).  
In addition, the USNO CCD Astrograph observations are affected by
remaining systematic errors on the same 10 to 20 mas level due to
the poor charge transfer efficiency of the CCD detector used
\citep{ucac3px}.  The size of the astrograph field-of-view is
1 by 1 degree.
From both sources, the Tycho-2 and the astrograph observations,
we have possible systematic positional errors on the 10 to 20 mas level,
while there is no indication of remaining systematic errors 
in our deep CCD imaging data on the 10 mas or larger level.

In order to probe for small-scale correlations of optical$-$radio
position differences we produced a list of all nearest neighbor
cases, up to about a 1.2 degree separation, which are shown in
Table 10.  All sources which have optical$-$radio position
differences exceeding about 1.5 times their total, formal error
in either coordinate are marked in the last column referring to
the figure number of a zoomed-in sky plot showing the location
of those sources, the optical$-$radio position difference vectors,
and the total, formal error (circles).  
In Figs.~10, 11, and 12, data from the UCAC2-type
reductions are shown in red, while those of the UCAC4-type are
in blue.  Note, for many sources the errors for both types of
reductions for a given source and observing run are the same
(rounded to nearest mas),
and the blue color overwrites a red circle in those cases.

Fig.~10 shows an area containing 3 sources (0403$-$132, 0405$-$123, and 
0406$-$127 from left to right) all within about 1 degree on the sky.
For all sources more than one observation was performed at different 
epochs and results from all such observing runs are shown by multiple 
vectors and error circles.  
Individual vectors per source show consistent results independent
of the UCAC2 or UCAC4 type reduction and observing run.
There is no reason be believe that all optical centers of emissions
are offset from the radio centers by about the same amount and direction
for those 3 different QSOs.  However, the observed optical$-$radio
position offsets of 0405$-$123, and 0406$-$127 are the same (within
the errors) and those of 0403$-$132 are of similar amplitude but
with inverted direction.
Note that different Astrographic Catalogue plates (shown as
large squares in Fig.~10) cover this area of sky, with source
0403$-$132 being on the northern edge of the southern plate
(with a plate center near $-14^{\circ}$) while the 2 other sources 
are just south of the more northern plate center.
Remaining systematic errors from a coma term 
in the reductions of those plates could explain the change of sign 
around declination $-13^{\circ}$.  

Figs.~11 and 12 show the other 2 cases of neighboring sources 
in our sample with significant optical$-$radio position differences.
The most northerly source in Fig.~12 does not provide any insight
because the errors are much larger than the optical$-$radio
position difference.  The other 2 sources in Fig.~12 show some large
optical$-$radio position differences along similar directions,
while the data from Fig.~11 do not.
The amplitude of optical$-$radio position offsets
seen in our data is well within the expected, remaining, local 
systematic errors of Tycho-2 reference stars at our observing epoch
(see above).

Looking at all results (Figs.~8 to 12) indicates that remaining
systematic errors in the reference stars (Tycho-2 proper motions
via Astrographic Catalogue, and possibly also UCAC data) could very 
well explain at least some of the larger than expected (from all 
known formal errors) optical$-$radio position differences.
Using spherical harmonics analysis on the entire southern 
hemisphere is unlikely to give more insight into this problem.
Large-scale zonal patterns might be revealed; however,
apparently small-scale variations of systematic errors over just 
1 to 2 degrees seem to play a major role and the number of sources 
is insufficient to 
characterize errors on such small scales with spherical harmonics.

\subsection{Physical offset of optical and radio centers of emissions}

Are there other explanations which might explain part of the
larger than expected optical$-$radio position differences than
issues with optical positions of reference stars?
The answer is yes, there are indications for actual physical position
offsets between radio centers and optical counterpart centers.

The first indication comes from sources with very large optical$-$radio
position differences \citep{iau2012}.  Source 0648$-$165 was observed
twice with formal errors of 30 and 70 mas per coordinate, respectively,
while the optical$-$radio offsets are consistently about 300 to 400 mas 
per coordinate.
This is much larger than could be explained from remaining systematic
errors in some reference stars.

Another example is 1345+125, a Seyfert galaxy with visible host
galaxy and 2 nuclei, separated by about 2 arcsec \citep{odea}, 
resulting in our observed optical$-$radio position offset of 
about 1.3 arcsec from data taken in about 1.5 arcsec seeing
which did not resolve the components.
A similar blended image example is 1730$-$130 which was not
resolved in one of our observing runs and resulted in over 400 mas
optical$-$radio position offset, while special point spread function
(PSF) handling of data from another observing run with better seeing
allowed the blended image to be resolved, resulting in an
optical position coinciding with the radio source position 
within the formal errors of about 30 mas.

Other investigations also find such outlier cases (see below),
which could be explained by some abnormality in the optical
structure (blended images, optical galaxy core not coincident
with radio peak etc.).  The question is, do we only have
to exclude such sources and arrive at a ``clean"
set of sources with no physical position offsets between centers
of radio and optical images, after which all the remaining excess 
in optical$-$radio position offsets can be explained by reference
star issues?  The answer is likely no, as we will see next.

The ratios of our optical$-$radio
position differences and their formal errors (per coordinate)
were used for the following, excluding 6$\sigma$ and larger outliers.
Fig.~13 shows the RMS scatter of those ratios (binned by 16 sources)
as a function of redshift, while Fig.~14 shows the same as a function
of X-band radio structure index.
There is an obvious trend toward smaller optical$-$radio position
differences for larger redshifts and smaller radio structure index.
This somewhat surprising result confirms earlier findings of this
nature \citep{dasilva}.
This effect can not be explained by systematic errors in reference 
star catalogs, and contradicts the assumption that optical and
radio centers of emissions are coinciding on the mas level independent
of radio structure or distance (redshift) of the sources.
However, the radio structure is only on the (sub)mas level while the 
observed unexplained excess in the optical$-$radio 
position offsets are 1 to 2 orders of magnitude larger.
Similar to the connection between the mass of the black hole
in the center of galaxies and the overall size of the galaxy,
there seems to be some connection between small-scale radio
structure and larger scale optical$-$radio position offsets.

\section{Comparison to other projects}

A comparison to \citep{camargo} is not made here due to the small
number of sources (22) in common and relatively high optical position 
errors (80 mas).
A recent paper \citep{orosz} published optical positions of over
1200 ICRF2 counterparts based on SDSS observations.
However, no source in common was found between these (mostly northern
hemisphere) and our observations (mostly southern hemisphere and original 
ICRF sources).

A total of 281 sources of the 300 sources in the Rio survey \citep{riosurvey}
are in common with our observations.  The unweighted, mean position
differences of our optical observations to those of the Rio survey
are $-3.3$ and $-8.1$ mas for the RA and Dec component, respectively
when using the UCAC2-type reductions.  The RMS scatter for these
differences (excluding individual position differences larger than
200 mas) are 43 mas per coordinate.  Using the UCAC4-type reductions
the mean position differences with the Rio survey are $-1.5$ and $-6.5$ mas
for RA and Dec, respectively with an RMS scatter of 45 mas per coordinate.

Fig.~15 shows the optical$-$radio position differences as observed
by the Rio survey versus that of our observations.
As expected there is some correlation, both sets of data point to
similar optical$-$radio position differences because both are based
ultimately on the Tycho-2 optical reference frame with some sort of
UCAC data for intermediate reference stars, and the deep CCD imaging
has the same resolution and similar mean epoch of observation in both
cases.
However, this comparison does not help us to distinguish between
reference stars systematic errors and possible physical position offsets
between the radio and optical centers of emissions.

As part of the Hipparcos to ICRF coordinate system link the separation
between pairs of selected QSOs and nearby Hipparcos stars was measured
with the HST Fine Guidance Sensors (FGS) \citep{hstfgs}.
System orientation and rotation (spin) angles (6 parameters) were 
determined with an accuracy of about 2 mas (2.5 mas/yr) from observations 
of about 40 pairs.  Elaborate calibrations 
to the FGS data were applied related to interferometer null, coordinate 
system drift, and changes in field angle distortions.
Nevertheless the input variances of both the 
formal errors of Hipparcos proper motions as well as the formal errors 
of the FGS observations had to be scaled by a factor of 4 to match the
error budget of the solution. 
Post-fit residuals on the 10 mas level were observed, significantly
larger than the expected formal errors.
An individual target, for some yet unknown reason, had to be excluded 
which changed the entire solution by over 10 mas/yr for the rotation 
around the $z$-axis. 
 
Another interesting clue comes from comparing the number of
``outliers" to the sample size of various investigations when
considering the optical observation precision.
The comparison of SDSS optical positions with ICRF2 radio sources
\citep{orosz} finds 51 ``outlier" sources (over 3$\sigma$ of formal
errors) out of 1297 sources total, i.e.~about 4\% of the
optical position data at about the 60 mas precision level.
Both this paper and our investigation explicitly list all the outliers 
found, while
the Rio survey \citep{riosurvey} of 300 sources was cleaned up
before publication and does not list any sources with optical$-$radio
position offsets over 3$\sigma.$  The precision of the Rio survey
is about 52 mas per coordinate when RMS combining 
the source centroid errors and zero-point errors from the 
``plate" reduction process.

For our UCAC2-based data we find 145 observations out of a total of 666
where at least 1 coordinate has an optical$-$radio position offset of
3$\sigma$ or more of the total, estimated random position error.
These belong to 104 unique sources out of 371 sources, thus about
28\% are ``outliers" under this definition.
For the UCAC4-based results we see similar numbers: 156 observations
over 3$\sigma$ out of 682 total, for 113 unique sources out of 392,
or 29\%.
If we attribute some of the unexplained errors to reference
star issues and assume the above mentioned best fit Gaussian
curve to the observed optical$-$radio positions (thus increasing
our error budget by about a factor of 4/3) that leaves us with 
about 63 ``outlier" sources (Table 9) out of a total of 371, 
which is 17\%.

\subsection{Are all sources ``problem cases" ?}

Thus between our data and the SDSS/ICRF2 investigation we see an
increase in the percentage of ``outliers" by about a factor of 20/4 = 5
for an increase in survey precision by about a factor of 2
(60 mas of SDSS/ICRF2 data versus 30 mas of our observations).
If this trend continues (ratio of percent increasing proportional to
ratio of survey precision) we would reach a 100\% ``outlier" level for
another precision factor of 2, thus for a survey of about 15 mas
precision per coordinate.
Note that many of our ``outlier" sources are among the higher
than average precision observations, i.e.~drawn from about 10 mas
1$\sigma$ position error samples. A 3$\sigma$ on that would be 
about 30 mas.
A survey similar to ours but 2 times more precise would thus show
most sources to be ``outliers" on the 15 mas level.

This leads us to the DARN hyphothesis (detrimental astrophysical
random noise) of optical$-$radio position differences:
For most ICRF-type sources the optical and radio centers of
emission are randomly offset by order of $\approx$ 10 mas due to
inherent astrophysical source structure effects (when imaged in 
the optical with a resolution too low to actually see individual
structure, and thus the direction and amplitude of possible
offsets).

Is this possible?  At typical distances of these ICRF quasars
(z = 0.5 to 4) the standard cosmological model predicts an
angular distance of about 1 Gpc within a factor of 2 
\citep{AGNbook},
thus a separation of features of 1 kpc corresponds to about 200 mas.
The brightness of host galaxies at optical wavelengths is typically
10\% of that of the dominant core of such AGNs, see for
example an imaging study with HST of z = 2 QSOs \citep{qsoimage}.
There is no reason to believe that the optical light distribution
of such a host galaxy is perfectly (on percent-level) symmetric
and centered on the radio core position. 
A near-IR adaptive optics study finds about 30\% of QSO host galaxies
disturbed \citep{AOimage}, which is 
not surprising considering the cause of AGNs related to merging galaxies.
The typical radius of AGN host galaxies is about 10 kpc \citep{AGNbook}.
Variations of optical light distributions (spiral arms, dust, asymmetry or
small offsets of halos in elliptical galaxies) could be on the kpc scale.
Assuming the combined center of optical light of this about 10 kpc
radius structure is offset by a mere 0.5 kpc
from the radio emission center (QSO black hole core), this translates
to an angular offset of 100 mas (see above).  With about 10\% of
the light contribution by the host galaxy the optical centroid of
the combined light from host galaxy and QSO would be shifted by
about 10 mas from the radio position when the optical structure 
can not be resolved (would need about 10 mas resolution and high
dynamic range).  
This seems like a possible if not even plausible scenario, at
least for ``disturbed" galaxies, while the effect will be less
for more symmetric, elliptical galaxies. 

We do not know the distribution of the true optical$-$radio
centers of emission offsets.  
A significant fraction of sources could have offsets much smaller
than 10 mas - we just don't know at this time. 
Gaia will be able to answer this.
It is doubtful that the problem of non-coincidence of optical
and radio centers of emissions can be mitigated by using better 
fitting algorithms, see e.g.~a discussion in \citep{quasometry}, 
if the disturbance causing the optical position offset is within
the angular resolution limit of the optical data.

A general increase in optical$-$radio position offsets noise on
the 10 mas level could of course come from remaining systematic
errors in reference star positions.  The amount of this error
amplitude is well withing the expected, remaining systematic
errors from reference stars, and if indeed degree-scale variations
of these errors on the sky exist, as indicated from the
above discussion, they would appear as random noise in our data.
However, that would not explain the correlations of the observed
optical$-$radio position offsets as a function of radio structure
index or redshift, nor the larger than expected errors in the
HST investigation for the Hipparcos frame alignmnet.  
Also, clear cases with astrophysically caused
offsets of optical and radio centers of emissions are seen on the
100 to 1000 mas level for a few (often nearby) sources. 
There is no reason to believe that the relative number of those
sources will decline when looking at smaller observed offsets
- rather the contrary is to be expected.
The few sources with over 100 mas astrophysically caused
optical$-$radio position offset might very well be the
tip of the iceberg.

For any radio to optical reference frame link effort, the errors
to be considered for a single object would be the RMS sum of
3 components: the observed position errors in the radio, 
the corresponding errors in the optical data and DARN.  
Most ICRF radio position errors are on the order of 1 mas or less, 
and the optical position errors expected from the upcoming Gaia mission
are on a similar level, which means the optical$-$radio reference
frame link of these 2 systems would be dominated by DARN.
Optical position shifts caused by variability of quasars linked
to their inner structure \citep{popovic} are likely on smaller
angular scales than DARN and thus might be overshadowed by DARN.

\subsection{Next steps}

In order to be better able to separate effects introduced by
systematic errors of reference stars and possible, astrophysical
non-coincidence of the radio and optical centers of emission,
another observing run at the 0.9 m CTIO telescope was performed
in March 2013.
Systematic errors in the reference stars are mostly expected to
come from proper motion errors.  For the data presented in this
investigation we had to bridge about 2000 $-$ 1991 = 9 years of
Tycho-2 proper motions, while the new observations have to bridge
2013 $-$ 1991 = 22 years.  This should lead to a significant
increase in the optical$-$radio position differences if indeed
reference star issues are dominant.  
Any optical structure induced optical$-$radio position offsets
of our sources are expected to stay the same because patterns
in the host galaxy can only change over much longer timescales.
Of course the random errors from proper motion errors also
increase and secondary reference stars were observed with a
different telescope (URAT, 
http://www.usno.navy.mil/USNO/astrometry/optical-IR-prod/urat)
this time, making the interpretation of the data more complex.
Possible short-term photometric variability of AGN cores can
shift the observed optical center of light as well, e.g.
\citep{taris}, complicating the situation further.
Results from the new observations will be published in an
upcoming paper.

Conclusive results in this matter will come from Gaia observations.
The internal errors of optical positions are sub-mas even for the
faint QSO targets.  However, if DARN is confirmed, the hoped for
accuracy of the alignment of the Gaia to the ICRF frame can not
be achieved by an order of magnitude or more.
Selecting a ``clean" sample e.g. by monitoring optical variability
\citep{gaia} 
will not help when a significant and unknown optical$-$radio position 
offset exists on top of possible image shift variations.
Small time-scale variations will be related to the AGN core,
while a general position offset is caused by the asymmetry of
the host galaxy's light distribution.

A better link of the Gaia and ICRF reference frame will come
from observing more sources.  Finding more optically bright QSOs
among radio sources \citep{obrs2} is an option.
Deep ground-based imaging can
determine positions of optically faint ICRF2 sources
using Gaia reference stars, even if the targets are too faint
to be observed with Gaia.  Finding more extragalactic radio
sources with optical counterparts will help, even if position
errors are relatively large (10 mas level).

\section{Conclusions}

Based on the results and discussions presented in this paper
we can draw the following conclusions:

\begin{enumerate}
\item This investigation is the most precise of its kind so far with a
  formal mean error of 25 mas per coordinate for the optical positions
  of ICRF counterparts. This was achieved from well calibrated astrometric
  observations with mostly the 0.9 m CTIO telescope in combination
  with contemporary astrograph observations providing intermediate bright
  reference stars on the Tycho-2 (Hipparcos) coordinate system (HCRS) 
  at the same epoch as the deep imaging data.

\item This sets an upper limit of about 25 mas (1$\sigma$) of possible
  non-coincidence of the optical and radio centers of emissions of
  typical ICRF quasars, with a few outliers already seen well exceeding 
  this level of positional precision.

\item Comparing UCAC2-type and UCAC4-type solutions, 
  use of UCAC2-type reference stars result in slightly smaller
  scatter of the optical$-$radio position differences, while use of
  UCAC4-type reference stars show smaller offsets to the
  radio reference frame coordinate axes.

\item Larger than expected (from the combined, total, formal errors)
  optical$-$radio position offsets are seen, which in part can 
  be explained by systematic errors in the reference stars, particularly
  errors in the Tycho-2 catalog changing on degree scales, likely due to 
  limitations in calibrating magnitude equations and coma terms in the 
  Astrographic Catalogue data.

\item Correlations of optical$-$radio position differences with
  redshift and radio structure index of sources, comparison of
  the number of ``outliers" from optical surveys of different
  precisions, and examining Hipparcos to ICRF link
  observations by HST FGS involving QSO-star pairs indicate other
  contributions besides systematic errors in reference star positions.
  These indications lead to the hypothesis of a true, detrimental, 
  astrophysical random noise (DARN) offset of radio and optical centers
  of emission on about the 10 mas level.  
  Higher optical accuracy data is needed 
  to conclusively prove this DARN hyphothesis; however, besides the 
  observational indications there is a possible astrophysical 
  explanation for such an error contribution on that level due to
  host galaxy structure at optical wavelengths.

\item It is estimated that
  Gaia observations will provide a rigid, instrumental, optical
  reference frame on the few micro-arcsecond level \citep{gaiasw}. 
  Its construction will not be affected by possible significant 
  optical$-$radio position offsets of individual ICRF sources.  
  However, the errors in alignment
  of the Gaia internal coordinate system to the ICRF will be affected
  by DARN, if it exists.  Although the radio positions as well as
  the Gaia observed optical positions of ICRF sources in common will
  have sub-mas precision, DARN will dominate the error budget. 
  If the mean optical$-$radio position offsets are indeed typically
  10 mas and there are about 600 link sources (2 coordinates each) the
  3 orientation angles of the Gaia frame each can only be determined
  to about 0.5 mas (1$\sigma$) precision with respect to the ICRF.  
  Absolute proper motions and parallaxes will not be compromised, 
  only absolute positions, if a rigid, instrumental Gaia system can
  be constructed without zonal errors.

\item  If DARN exists and a higher positional alignment of the
  Gaia and ICRF reference frames is desired, more objects in common
  with sufficient radio and optical flux will be needed, instead of
  concentrating on a ``clean" sample (i.e.~small radio structure
  index, low optical variability), because the error contribution 
  of an individual source for the radio to optical coordinate system 
  link will be dominated by that unknown, random optical position offset. 
  Ground-based observations of ICRF sources too faint to be seen with 
  Gaia directly could provide optical positions on the Gaia system using
  Gaia reference stars to its limiting magnitude in small fields of
  view, using large aperture, long-focus telescopes.
  Determining radio positions of more targets observed by Gaia is
  another option.  Radio or optical positions of those additional
  source observations would need to be only on or slightly better than 
  the position error level of DARN (maybe 5 to 10 mas), to provide meaningful, 
  additional data to improve the link of Gaia and ICRF reference frames.
  Monitoring programs to arrive at a ``clean" sample would be of lower
  priority than the shear number of ``reasonably good" sources with
  both optical and radio positional data.
  The stability of an optical or radio position (below the DARN level)
  would not matter because of the unknown significant position offset 
  between the centers of emission of optical and radio source.
\end{enumerate}


\acknowledgments

CTIO and KPNO are thanked for having granted observing time
and provided support for this project.
We would like to thank everyone involved in this over a decade long project,
particularly all observers beyond the authors (T.~Rafferty, E.~Holdenried,
and J.~Perez).  
Charlie Finch is thanked for running the dedicated astrograph observations 
through the UCAC4 pipeline for many of the fields.

Ralph Gaume is thanked for supporting this project as Head of
the Astrometry Department over the many years and under difficult
budget constraints.
National Optical Astronomy Observatories (NOAO) are acknowledged for
IRAF, Smithsonian Astrophysical Observatory for DS9 image display software,
and Tim Pearson at the California Institute of Technology for the 
{\em pgplot} software.
More information about the UCAC and follow-up projects is available at 
\url{www.usno.navy.mil/usno/astrometry/optical-IR-prod}.





\clearpage

\begin{figure}
\epsscale{1.00}
\includegraphics[angle=-90,scale=.34]{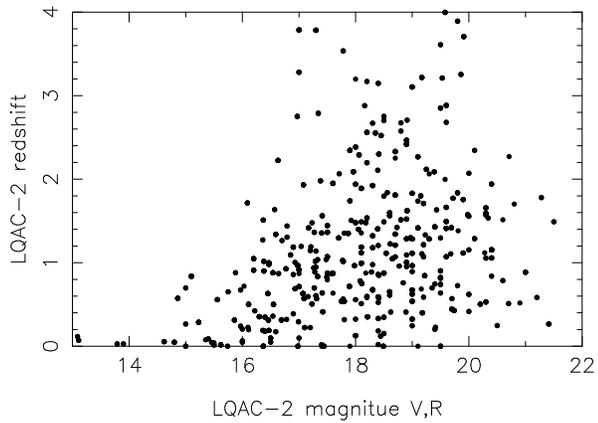}
\caption{Distribution of approximate magnitude (V if available, else R)
         and redshift of all sources in our investigation. 
         Data are taken from the LQAC-2 catalog.}
\end{figure}

\begin{figure}
\epsscale{1.00}
\includegraphics[angle=-90,scale=0.45]{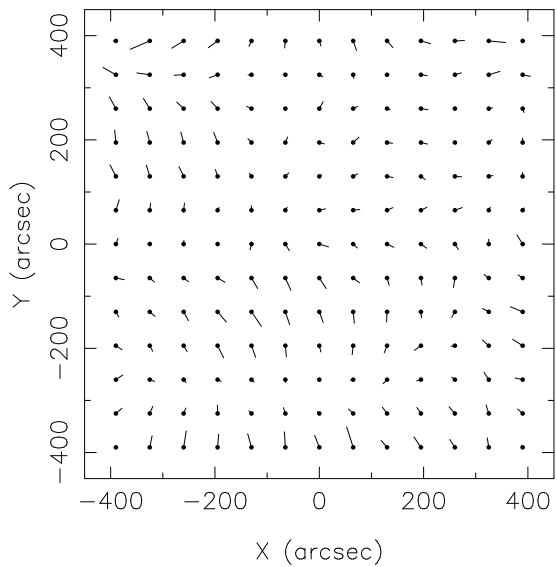}
\caption{Field distortion pattern of the CTIO 0.9 m telescope field of view.
         Smoothed data are shown from run ``q" and UCAC4-type reference 
         star residuals.  Vectors are scaled by a factor of 1000.}
\end{figure}

\begin{figure}
\epsscale{1.00}
\includegraphics[angle=-90,scale=0.33]{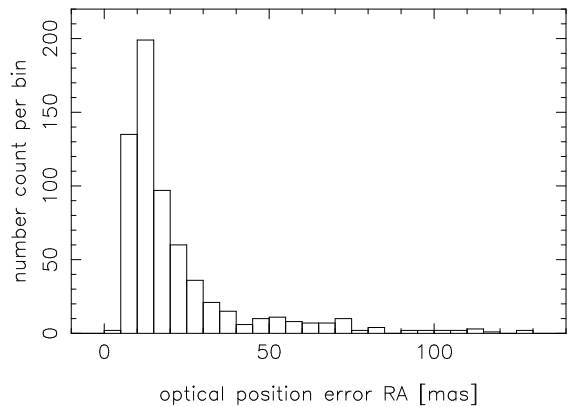}
\caption{Histogram of the total, random, mean optical position errors
         (per coordinate) from all exposures of an individual source and
         observing run.  
         These errors include $x,y$ fit errors, contribution from the 
         atmospheric turbulence and error propagation from reference stars.}
\end{figure}

\clearpage

\begin{figure}
\epsscale{1.00}
\includegraphics[angle=0,scale=0.45]{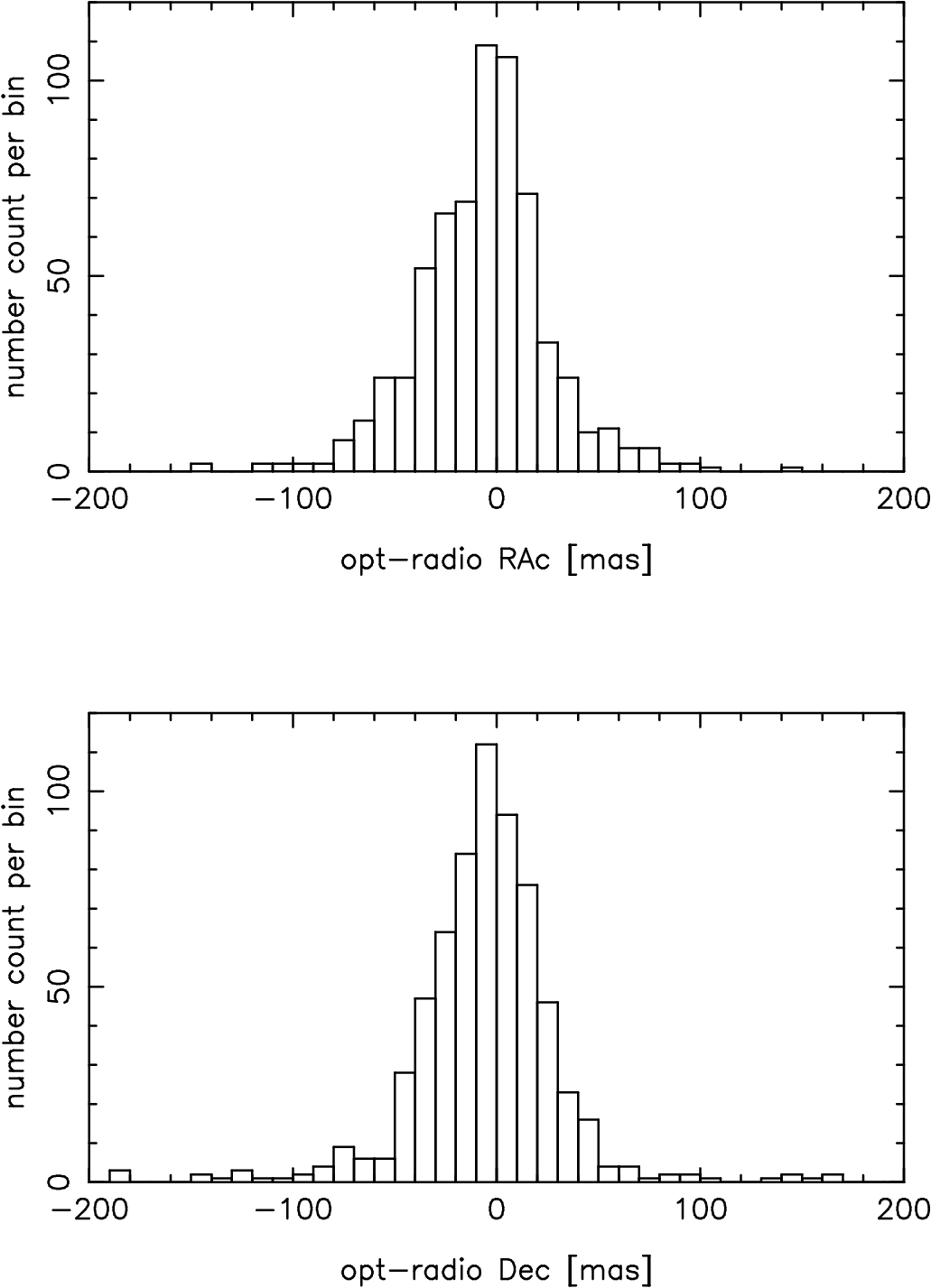}
\caption{Historgam of (optical$-$radio) position differences per
         coordinate of our UCAC2-type data.}
\end{figure}

\begin{figure}
\epsscale{1.00}
\includegraphics[angle=0,scale=0.45]{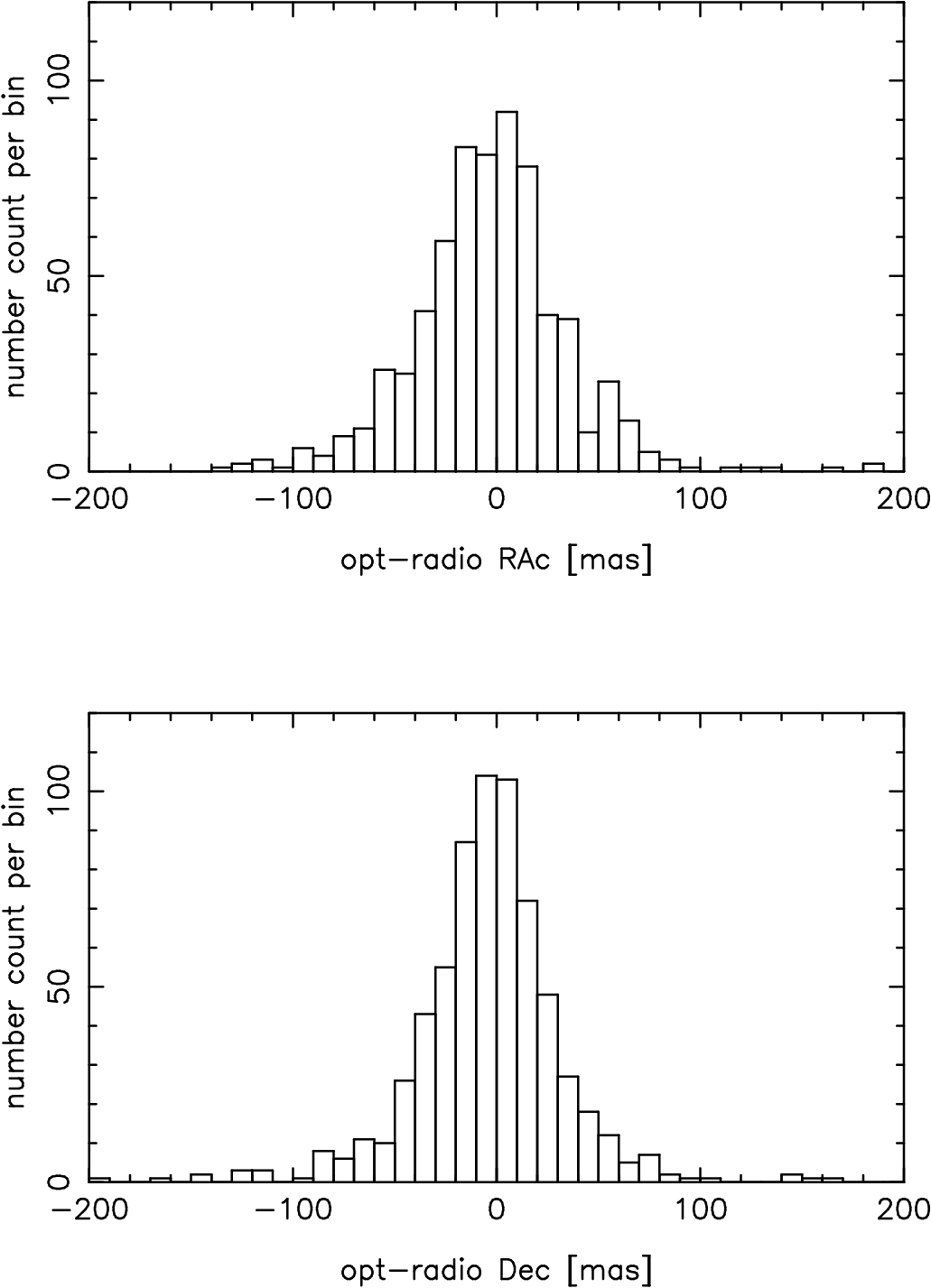}
\caption{Same as the previous figure but for UCAC4-type data.}
\end{figure}

\clearpage

\begin{figure}
\epsscale{1.00}
\includegraphics[angle=0,scale=0.45]{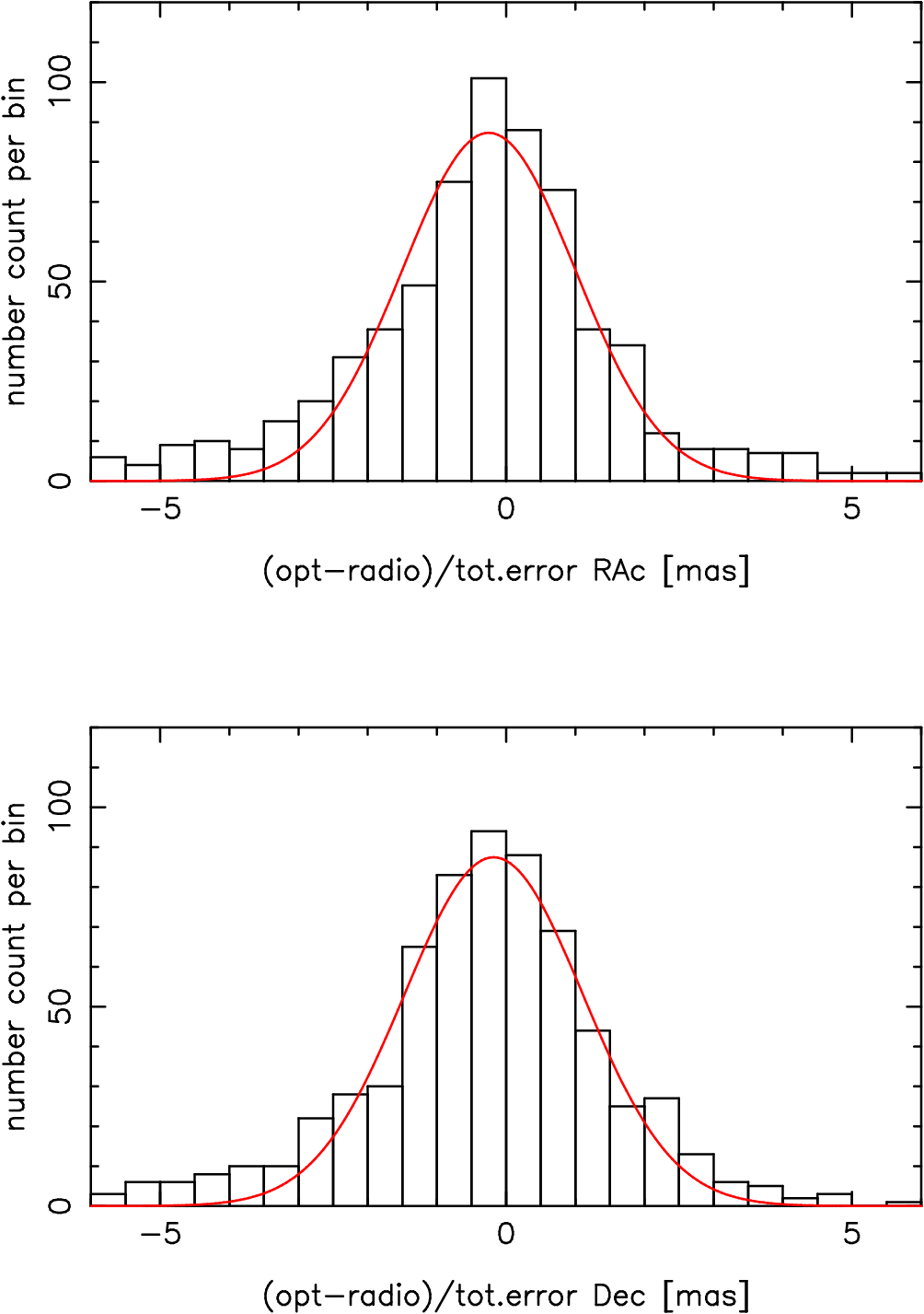}
\caption{Histogram of the scaled (optical$-$radio)/error position 
     differences per coordinate of our UCAC2-type data.
     The plot line represents the best fit normal distribution.}
\end{figure}

\begin{figure}
\epsscale{1.00}
\includegraphics[angle=0,scale=0.45]{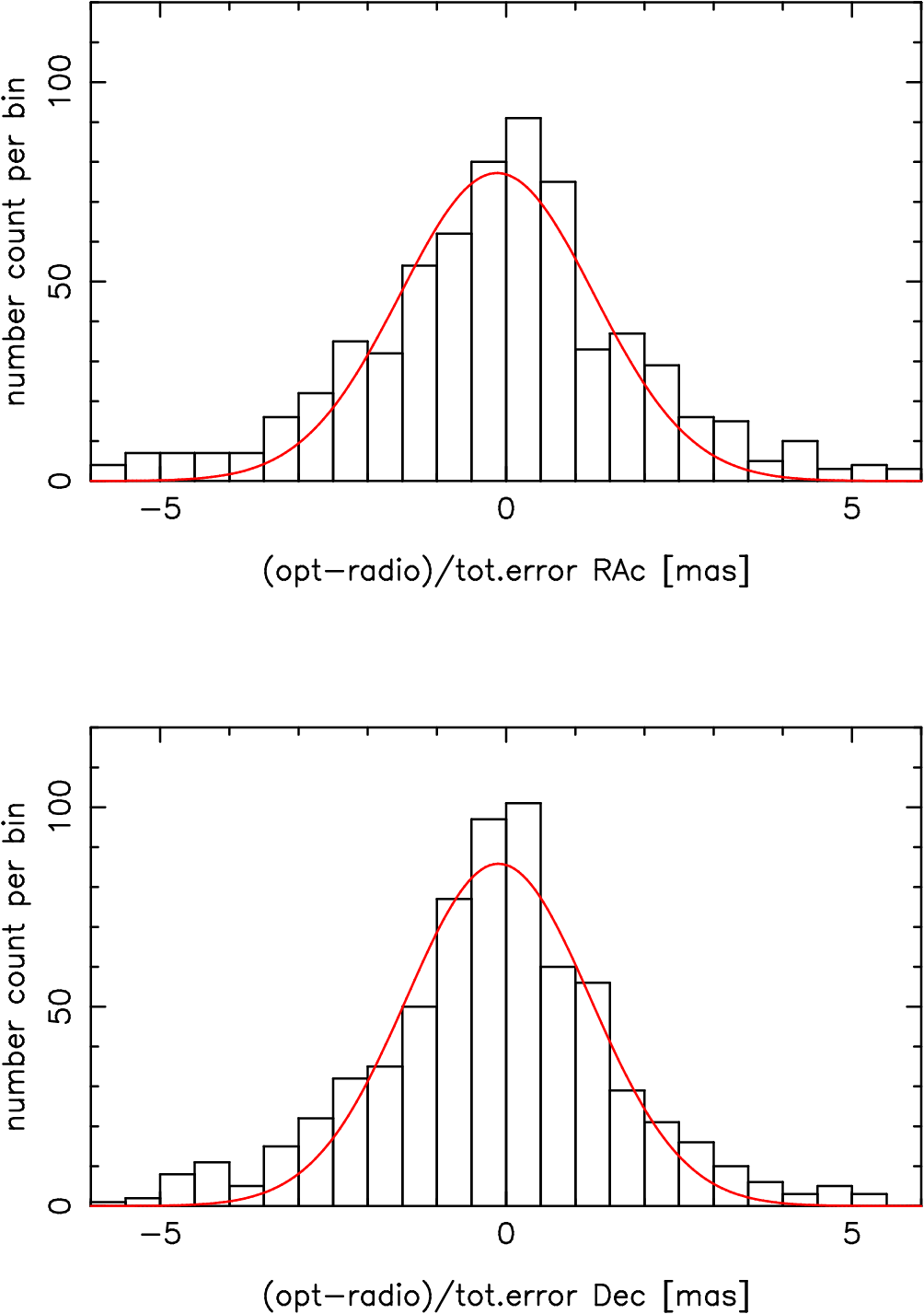}
\caption{Same as the previous figure but for UCAC4-type data.}
\end{figure}

\clearpage

\begin{figure}
\epsscale{1.00}
\includegraphics[angle=-90,scale=0.75]{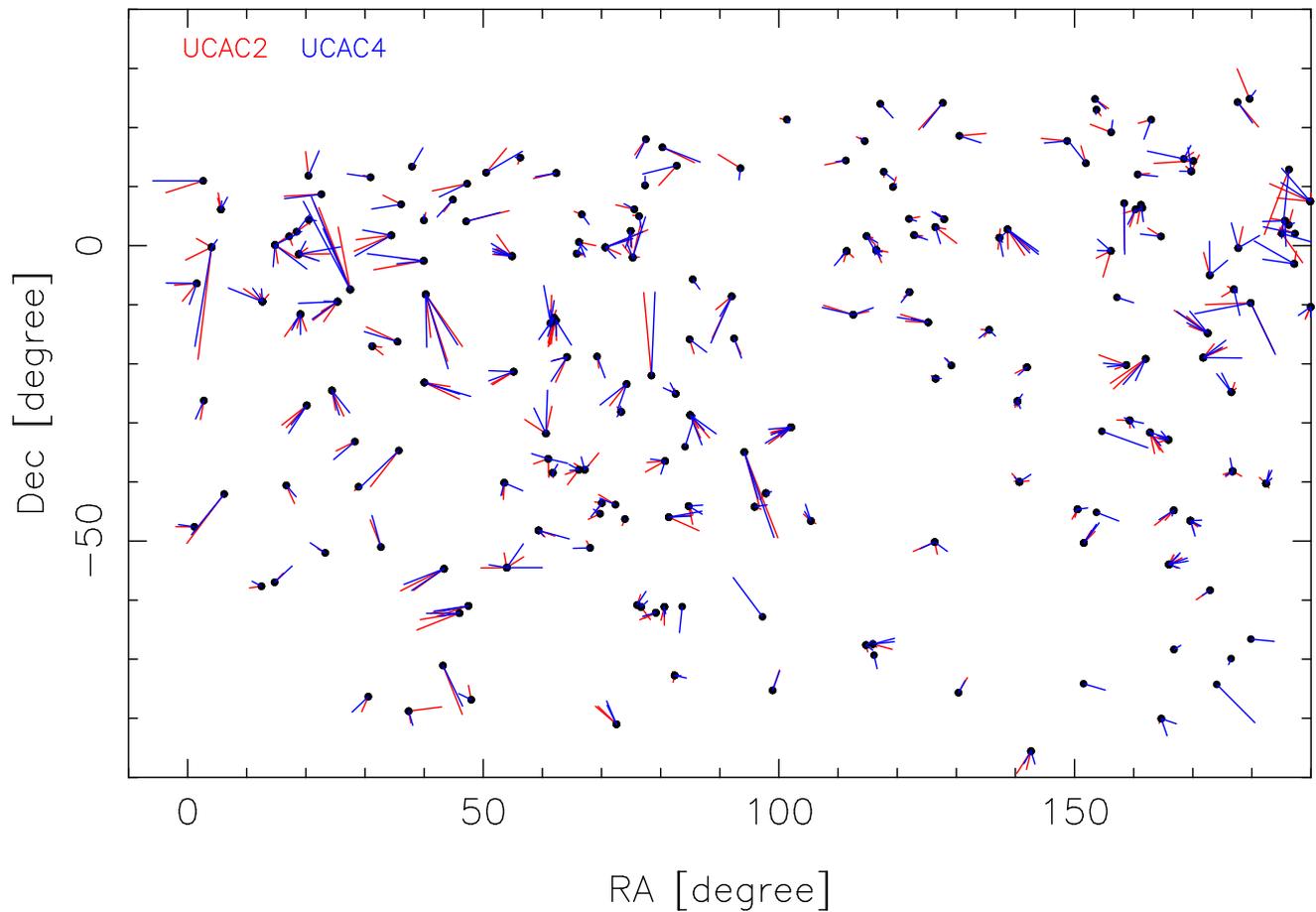}
\caption{Distribution of our sources on the sky with optical$-$radio 
   position difference vectors shown for each observation.
   Data from UCAC2 and UCAC4-type processing are shown in red and blue,
   respectively.  The scale of the vectors is 10 mas for 1 degree.
   Data for 0 th 12 hours RA (0 to 180 deg) are shown here.}
\end{figure}

\clearpage

\begin{figure}
\epsscale{1.00}
\includegraphics[angle=-90,scale=0.75]{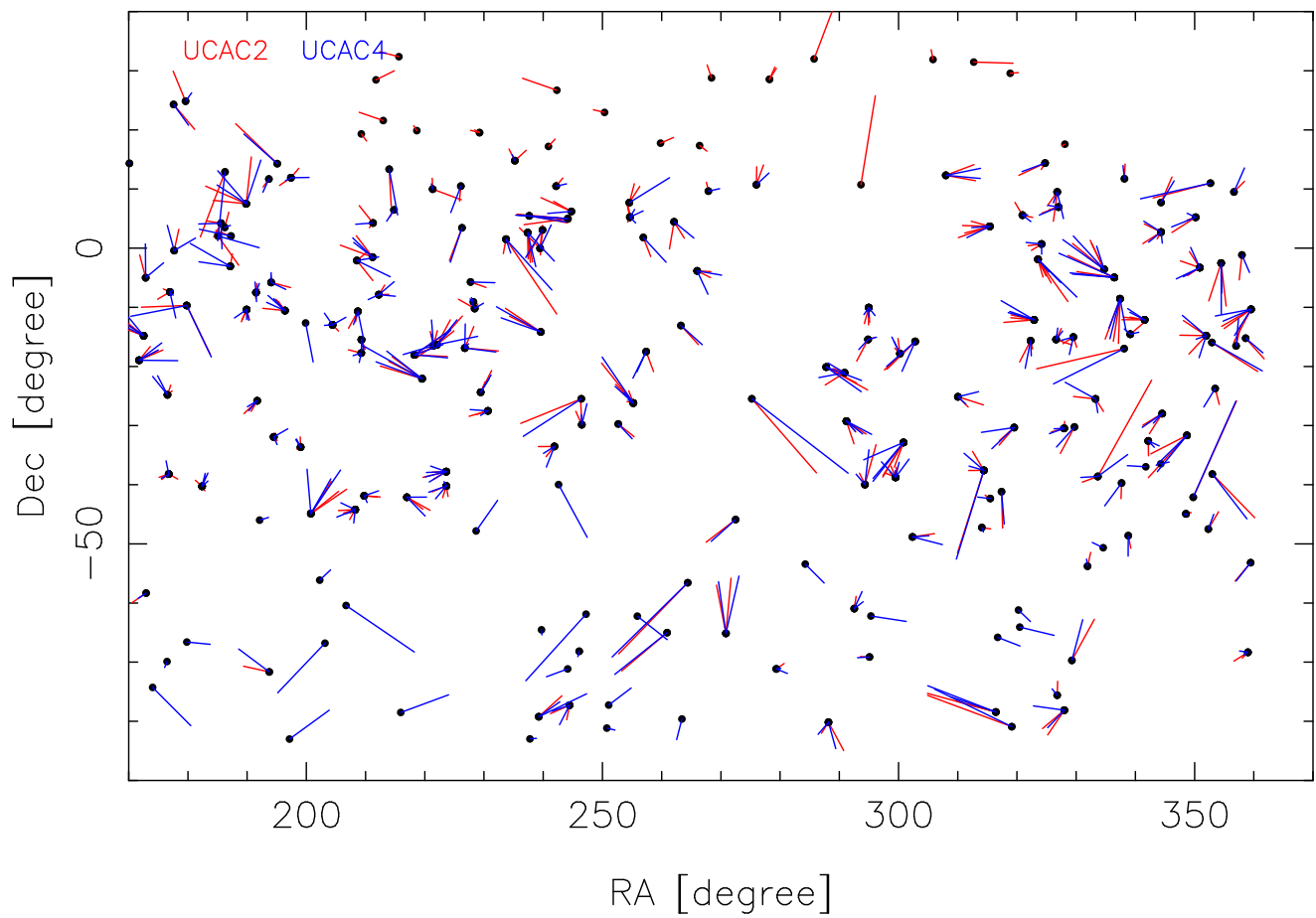}
\caption{The same as the previous figure but for RA = 12 to 24 hours
   (with some overlap).}
\end{figure}

\clearpage

\begin{figure}
\epsscale{1.00}
\includegraphics[angle=0,scale=0.50]{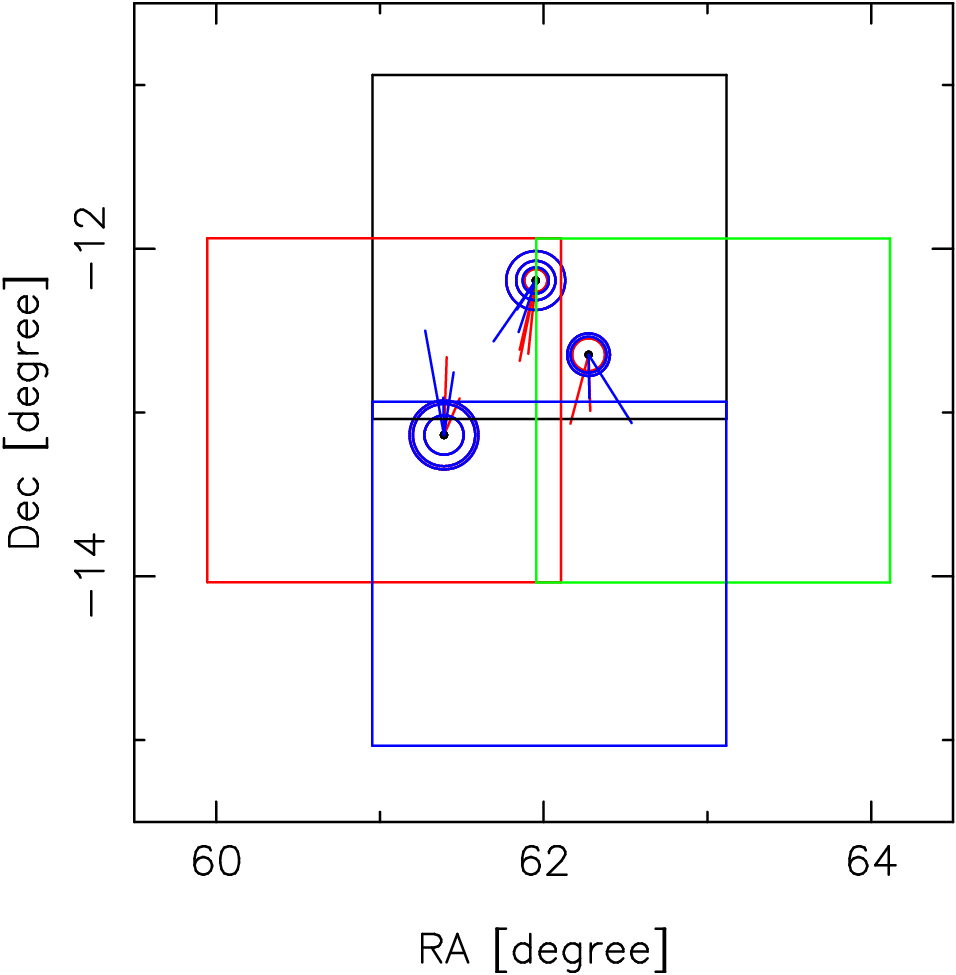}
\caption{A zoom in at the location of the 3 sources
   0403$-$132, 0405$-$123, and 0406$-$127 (from left to right)
   which are at redshifts 0.571, 0.574, and 1.563 respectively.
   Vectors show optical$-$radio position differences and circles
   the 1$\sigma$, total, formal errors.  The scale of the vectors
   and circles are 100 mas per 1 degree.  Results from multiple
   observing runs are shown.  UCAC2-type data are shown in red,
   while UCAC4-type data (overwriting UCAC2-type, if same values)
   are shown in blue.
   The squares indicate the Astrographic
   Catalogue plates used to derive the Tycho-2 proper motions.} 
\end{figure}

\begin{figure}
\epsscale{1.00}
\includegraphics[angle=0,scale=0.50]{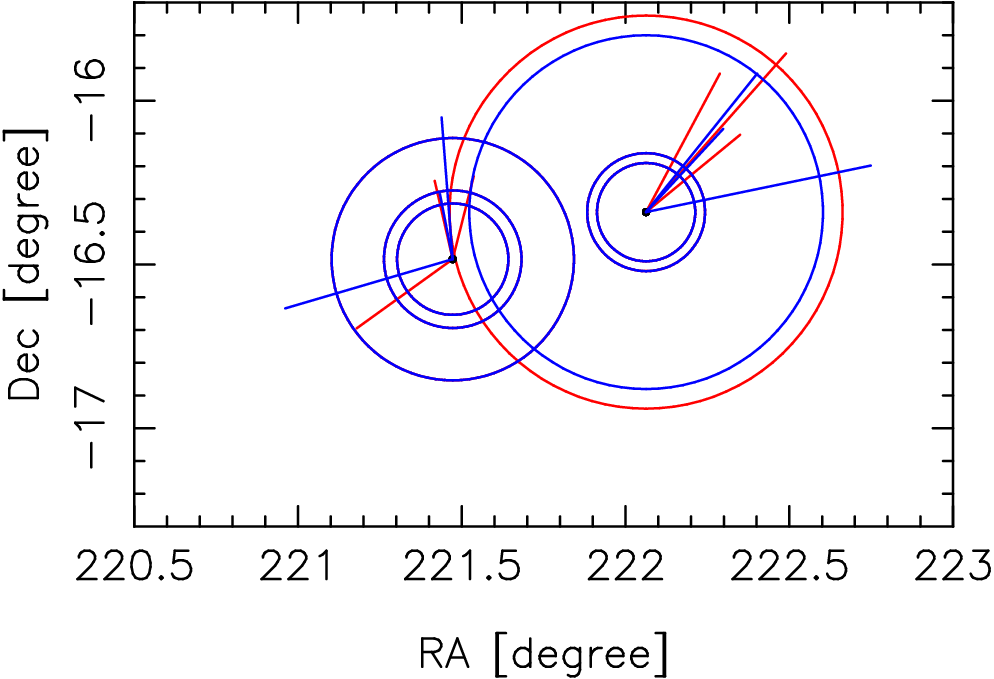}
\caption{Similar to the previous figure, a zoom in at the location 
   of the sources 1443$-$162 and 1445$-$161 which are at redshifts unknown 
   and 2.417 respectively.}
\end{figure}

\begin{figure}
\epsscale{1.00}
\includegraphics[angle=0,scale=0.50]{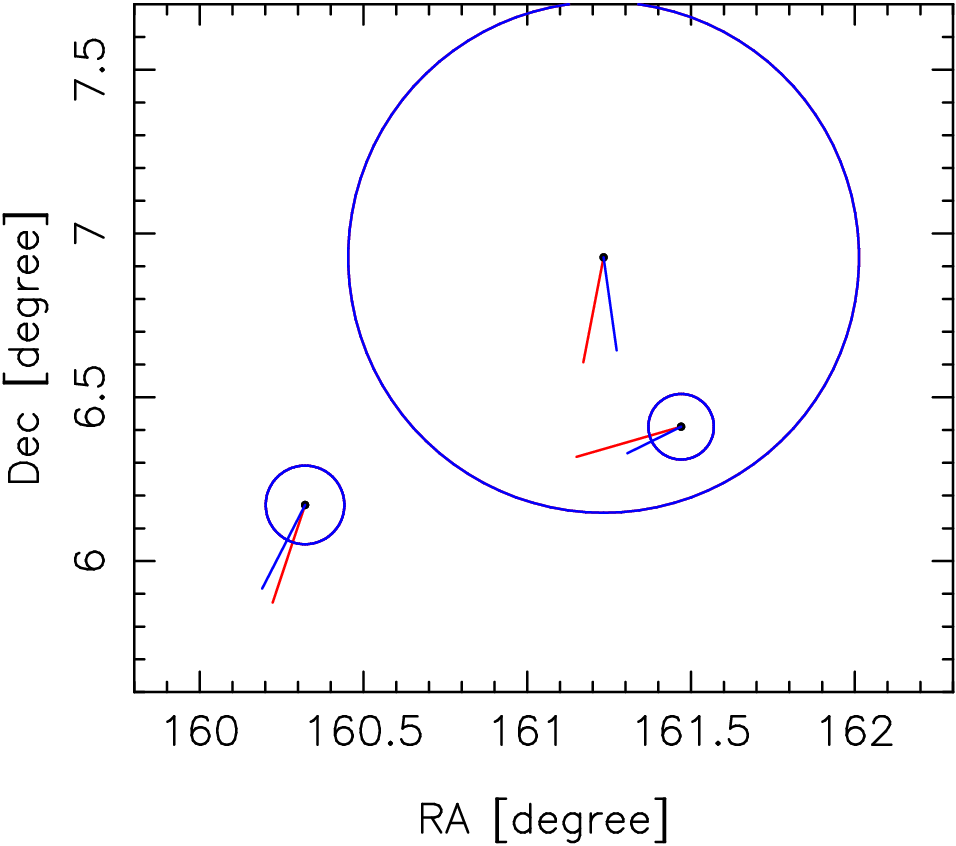}
\caption{Similar to the previous figure, a zoom in at the location of 
   the 3 sources 1038+064, 1042+071, and 1043+066 which are at redshifts  
   1.265, 0.698, and 1.507 respectively.}
\end{figure}

\clearpage

\begin{figure}
\epsscale{1.00}
\includegraphics[angle=0,scale=0.45]{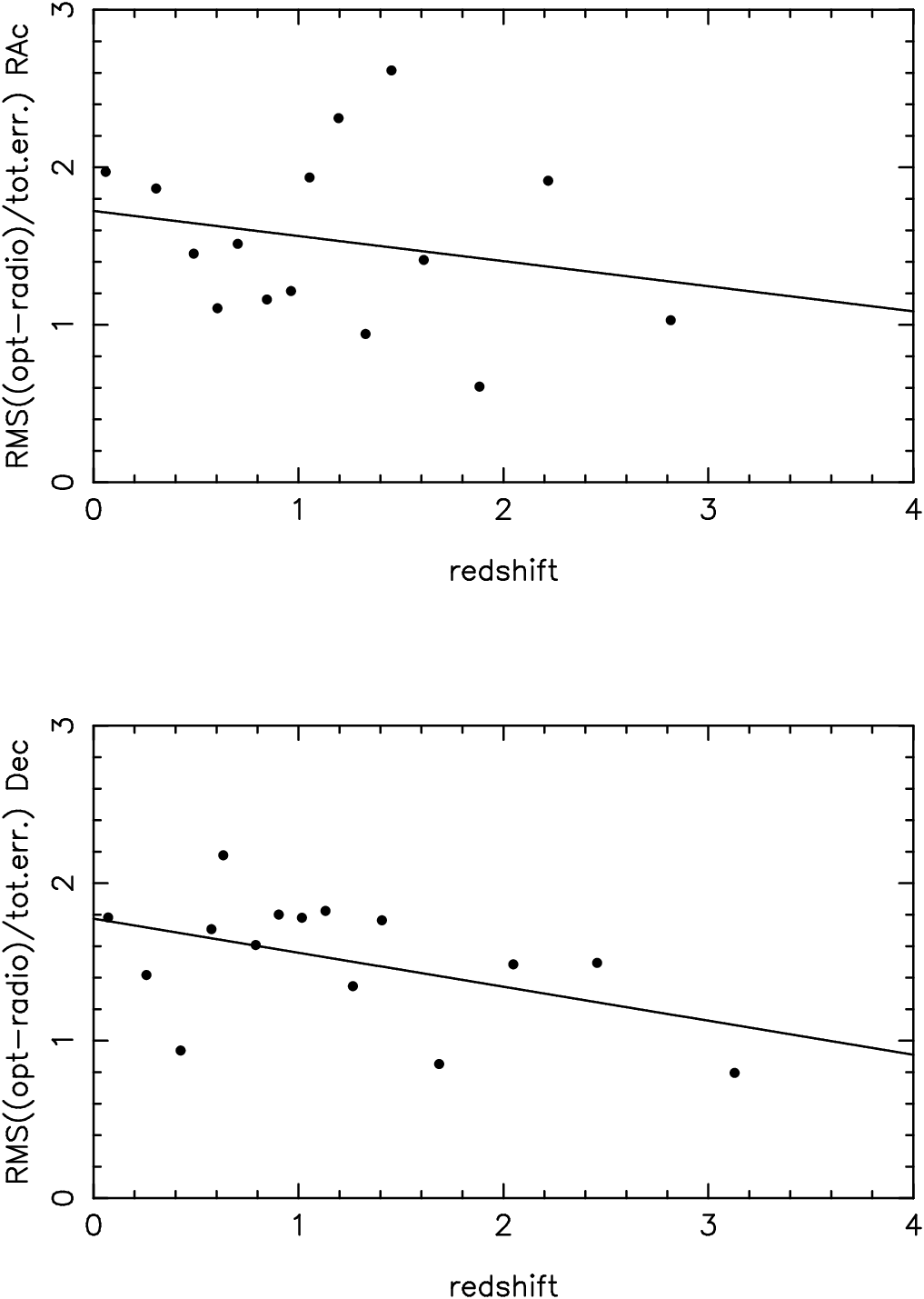}
\caption{Optical$-$radio position differences over total 
   formal error of individual sources (RMS over 16 offsets per dot)
   as a function of redshift, RA component on top, Dec at bottom,
   using UCAC2-based results.}  
\end{figure}

\begin{figure}
\epsscale{1.00}
\includegraphics[angle=0,scale=0.45]{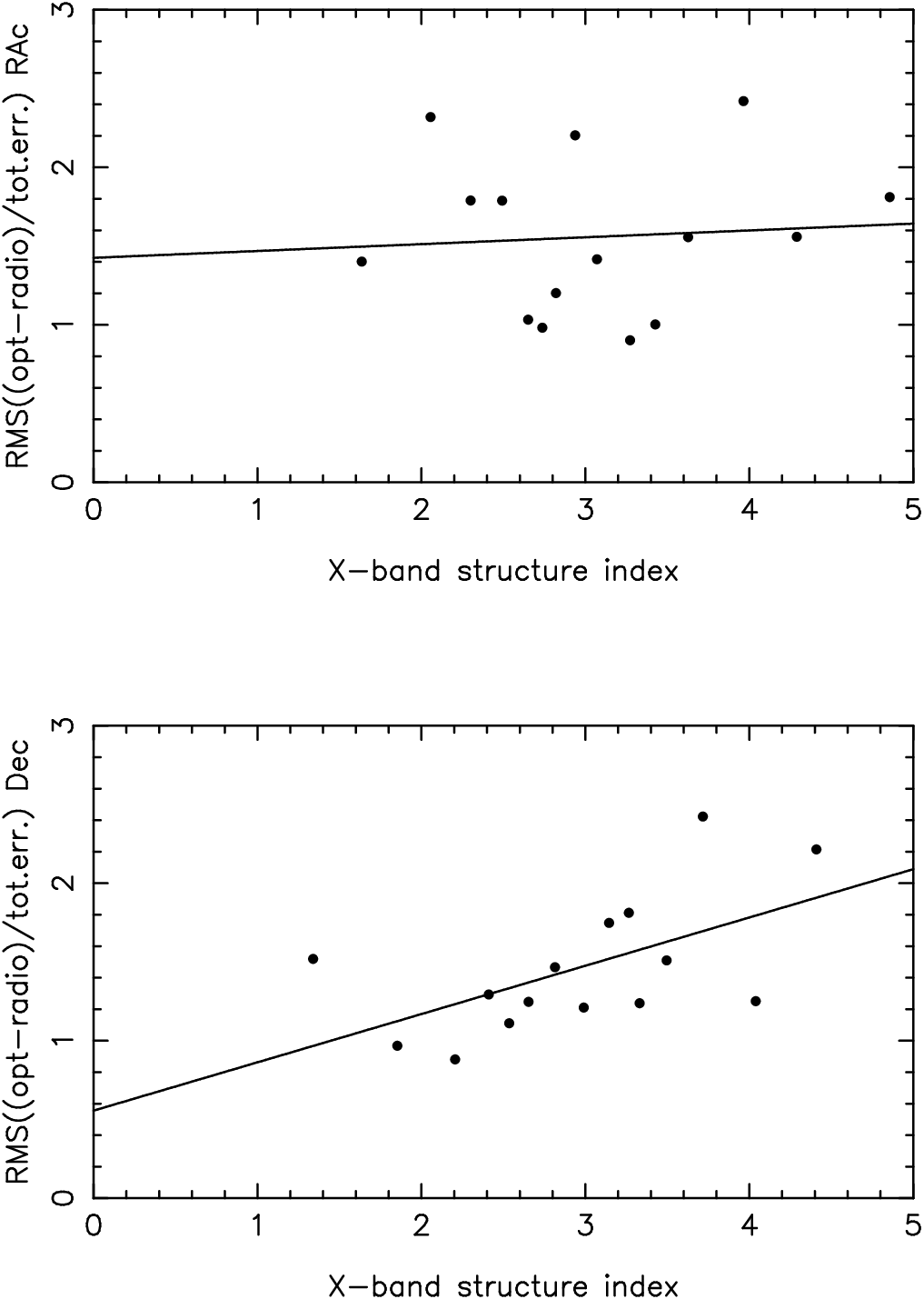} 
\caption{Same as previous figure but as a function of X-band
   radio structure index.}
\end{figure}

\begin{figure}
\epsscale{1.00}
\includegraphics[angle=0,scale=0.60]{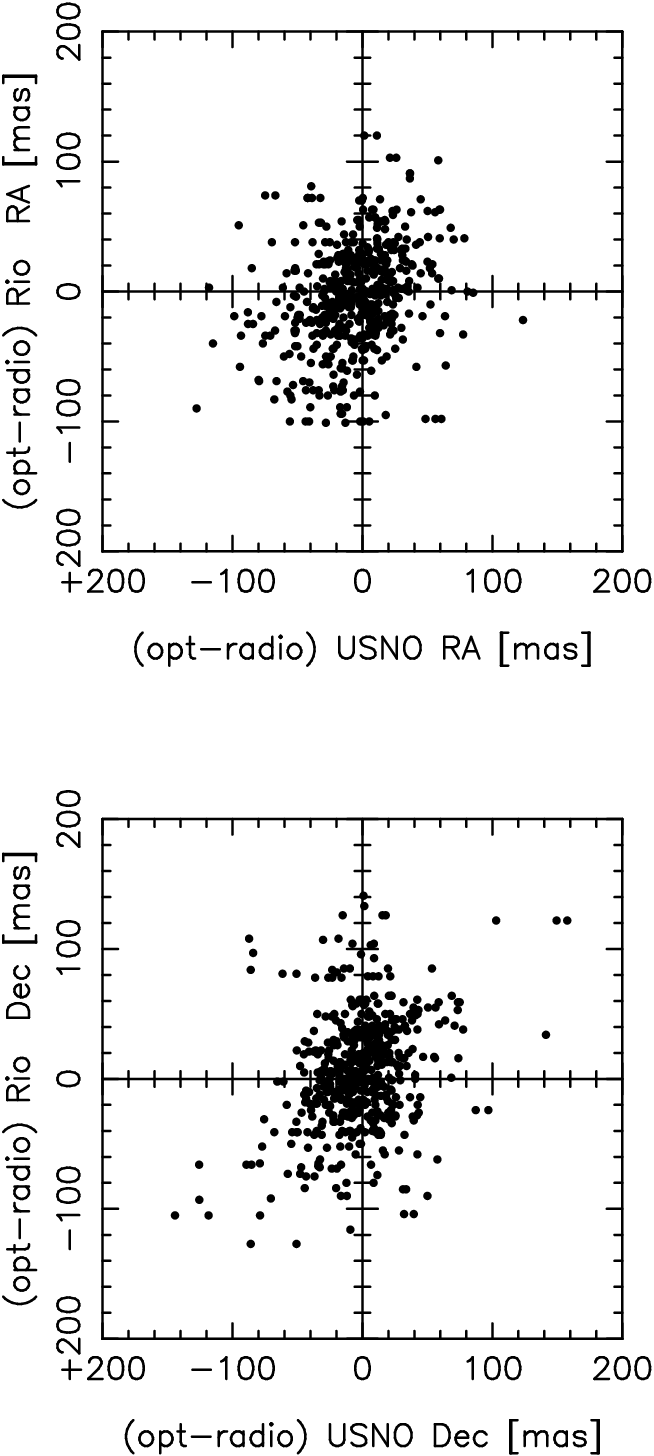}
\caption{Comparison of optical$-$radio position differences 
   of individual source observations of our investigation 
   versus the Rio survey data of the 281 sources in common.  
   The top figure shows the result for the RA component, 
   the bottom for Dec.}
\end{figure}




\clearpage

\begin{table}
\begin{center}
\caption{Overview of deep CCD imaging observing runs used for this 
  investigation.  Each run is identified with a letter.
  The telescope used is identified and the observing epoch is given.
  The last 3 columns give the number of observing nights, total number
  of ICRF source fields observed (including partial data and empty fields),
  and the initials of the observers (see acknowledgment).}
\vspace*{5mm}
\begin{tabular}{clllccc}
ID & telescope & \multicolumn{2}{c}{date}
   & n & s & obs \\
\tableline\tableline
o & CTIO 0.9 m & Dec&1997 &5& 37 & MZ, NZ \\
p & CTIO 0.9 m & Feb&1998 &5& 29 & MZ, NZ \\
q & CTIO 0.9 m & Jul&1998 &6& 46 & MZ, NZ \\
r & CTIO 0.9 m & Aug&1998 &6& 44 & MZ, NZ \\
s & CTIO 0.9 m & Jan&1999 &5& 56 & NZ, TR \\
t & CTIO 0.9 m & Mar&1999 &5& 56 & MZ, NZ, JP \\ 
u & CTIO 0.9 m & Jun&1999 &4& 61 & MZ, NZ \\
v & CTIO 0.9 m & Sep&1999 &6& 52 & TR \\
w & CTIO 0.9 m & Dec&1999 &5& 57 & NZ \\
x & CTIO 0.9 m & Mar&2000 &4& 77 & TR \\
y & CTIO 0.9 m & Jun&2000 &5& 51 & NZ \\
z & CTIO 0.9 m & Sep&2000 &5& 77 & TR \\
a & CTIO 0.9 m & Jan&2001 &5& 39 & EH \\
b & KPNO 2.1 m & Jun&2001 &5& 42 & MZ \\
J & CTIO 0.9 m & May&2004 &5& 66 & NZ \\ 
\tableline
\end{tabular}
\end{center}
\end{table}

\begin{table}
\begin{center}
\caption{Observing of ICRF source fields at the astrograph.}
\vspace*{5mm}
\begin{tabular}{lc}  
item  & description \\
\tableline\tableline
observing dates           & Jan 1998 to June 2003 \\
number of frames total    &    14,653          \\
number of frames rejected &   about 20 \%      \\
exposure times            &   30 and 150 sec   \\
typical number of frames  &  \\
\hspace{5mm} per run and source &      8 $-$ 16  \\
control of system.errors  & observe on both sides of pier \\
\hspace{5mm} plus observe & calibration fields frequently \\
astrograph imaging        & within 2 weeks of deep runs\\
\tableline
\end{tabular}
\end{center}
\end{table}

\begin{table}
\begin{center}
\caption{The USNO ``redlens'' astrograph and camera used for
         the secondary reference star observations.}
\vspace*{5mm}
\begin{tabular}{lrl}  
item  & \multicolumn{2}{r}{description} \\
\tableline\tableline
clear aperture             &  206  & mm   \\
focal length               & 2057  & mm   \\
number of lens elements    &   5   &     \\
corrected for bandpass     & 550$-$710 & nm \\
diameter usable field of view  & $\approx$ 9 & degree \\
active guiding with     & \multicolumn{2}{r}{ST4 at visual lens} \\
\hline
number of pixels        & 4095 x 4095 & \\
field of view           &  61 x 61  & arcmin \\
pixel size              &      9.0 & $\mu m$ \\
pixel scale             &    0.905 & arcsec/pixel \\
spectral bandpass used  & 579$-$642 & nm \\
limiting magnitude      & $\approx$ 16.0 & 2 min. \\
\tableline
\end{tabular}
\end{center}
\end{table}

\clearpage

\begin{table}
\begin{center}
\caption{Results from UCAC2-type reductions of ICRF optical counterparts.
  For a detailed explanation of columns see text.  Only example lines are
  shown here, the complete table with 666 lines is available as catalog
  I/325 from CDS.}
\vspace*{5mm}
\small
\begin{verbatim}
         1           2            3            4         5        6       7    8    9     10    11   12  13   14  15    16    17   18   19  20  21 22    23  24 25 26  27
    RAJ2000      DEJ2000        RAJ2000      DEJ2000S   dRAc    dDec    dist  eRA  eDE   sRA   sDE eRad Nccd  Nr Nrx   err   S/N  mag    z   T   S  U   name Run F Q  notes
 h  m  s        d  '   "         degree       degree     mas     mas     mas  mas  mas              mas                mas
00 04 35.6524 -47 36 19.603   0.07657011  -47.6054454   31.4   -31.4     0.3  13.  13.  -2.4   0.0  0.2   5   20   1  14.6    37 15.88 0.880 3 9.00 2 0002-478 r r g
00 06 13.8912 -06 23 35.366   0.10385868   -6.3931571   38.9   -24.5   -30.2  29.  29.  -0.8  -1.0  0.0   3   18   0  22.3    13 18.50 0.347 2 3.18 2 0003-066 v r g
00 06 13.8903 -06 23 35.338   0.10385843   -6.3931495   38.0   -37.9    -2.9  12.  12.  -3.2  -0.2  0.0   4   20   0  14.9    41 18.50 0.347 2 3.18 2 0003-066 z r l
00 10 31.0016 +10 58 29.484   0.17527823   10.9748568   66.0   -62.9   -19.9  13.  13.  -4.8  -1.5  0.1   4   15   3  28.1    11 15.40 0.089 1 1.82 2 0007+106 z r l  galaxy
00 11 01.2463 -26 12 33.409   0.18367953  -26.2092803   32.6    -5.8   -32.1  62.  62.  -0.1  -0.5  0.1   1   16   0  20.6    10 18.90 1.093 3 3.04 2 0008-264 w r g
00 16 11.0870 -00 15 12.635   0.26974640   -0.2535098  191.3   -22.7  -189.9 116. 116.  -0.2  -1.6  0.1   4   17   1  27.3     2 19.98 1.574 3 2.17 2 0013-005 v r f
00 16 11.0856 -00 15 12.467   0.26974601   -0.2534631   48.9   -43.8   -21.8  67.  67.  -0.7  -0.3  0.1   4   15   1  22.4     3 19.98 1.574 3 2.17 2 0013-005 y r f
00 16 11.0869 -00 15 12.431   0.26974635   -0.2534530   29.2   -25.4    14.5  44.  44.  -0.6   0.3  0.1   3   20   0  17.9    10 19.98 1.574 3 2.17 2 0013-005 z r g
00 22 32.4409 +06 08 04.297   0.37567803    6.1345269   28.2    -4.5    27.8  37.  37.  -0.1   0.8  0.1   4   21   1  22.2     7 19.50 0.000 2 1.78 2 0019+058 y r g
00 22 32.4408 +06 08 04.288   0.37567801    6.1345245   19.9    -5.6    19.1  19.  19.  -0.3   1.0  0.1   4   18   0  18.7    23 19.50 0.000 2 1.78 2 0019+058 z r g
00 24 42.9841 -42 02 04.032   0.41194003  -42.0344532  104.8   -63.2   -83.6  74.  74.  -0.9  -1.1  1.3   2   19   0  17.7     5  0.00 0.000 0 9.00 2 0022-423 r r f
\end{verbatim}
\normalsize
\end{center}
\end{table}

\begin{table}
\begin{center}
\caption{Results from UCAC4-type reductions of ICRF optical counterparts.
  For a detailed explanation of columns see text.  Only example lines are
  shown here, the complete table with 682 lines is available as catalog
  I/325 from CDS.}
\vspace*{5mm}
\small
\begin{verbatim}
         1           2            3            4         5        6       7    8    9     10    11   12  13   14  15    16    17   18   19  20  21 22    23  24 25 26  27
    RAJ2000      DEJ2000        RAJ2000      DEJ2000S   dRAc    dDec    dist  eRA  eDE   sRA   sDE eRad Nccd  Nr Nrx   err   S/N  mag    z   T   S  U   name Run F Q  notes
 h  m  s        d  '   "         degree       degree     mas     mas     mas  mas  mas              mas                mas
00 04 35.6528 -47 36 19.600   0.07657023  -47.6054444   27.4   -27.1     3.9  19.  19.  -1.4   0.2  0.2   2   20   0  21.8    37 15.88 0.880 3 9.00 4 0002-478 r r g
00 06 13.8920 -06 23 35.371   0.10385890   -6.3931585   37.5   -12.6   -35.3  29.  29.  -0.4  -1.2  0.0   3   21   0  28.0    13 18.50 0.347 2 3.18 4 0003-066 v r g
00 06 13.8894 -06 23 35.335   0.10385817   -6.3931485   51.8   -51.8     0.7  12.  12.  -4.3   0.1  0.0   4   26   0  23.2    41 18.50 0.347 2 3.18 4 0003-066 z r l
00 10 31.0001 +10 58 29.504   0.17527781   10.9748623   85.2   -85.2    -0.1  12.  12.  -7.1   0.0  0.1   4   29   0  38.0    11 15.40 0.089 1 1.82 4 0007+106 z r p  galaxy
00 11 01.2457 -26 12 33.408   0.18367936  -26.2092800   34.0   -14.0   -31.0  62.  62.  -0.2  -0.5  0.1   1   17   0  19.3    10 18.90 1.093 3 3.04 4 0008-264 w r g
00 16 11.0867 -00 15 12.613   0.26974630   -0.2535037  170.3   -28.1  -168.0 116. 116.  -0.2  -1.4  0.1   4   23   0  40.6     2 19.98 1.574 3 2.17 4 0013-005 v r f
00 16 11.0889 -00 15 12.441   0.26974691   -0.2534558    6.6     4.8     4.5  67.  67.   0.1   0.1  0.1   4   16   0  42.9     3 19.98 1.574 3 2.17 4 0013-005 y r f
00 16 11.0892 -00 15 12.433   0.26974700   -0.2534535   16.0     9.7    12.7  44.  44.   0.2   0.3  0.1   3   25   0  24.4    10 19.98 1.574 3 2.17 4 0013-005 z r g
00 22 32.4413 +06 08 04.284   0.37567814    6.1345232   14.5     1.4    14.4  37.  37.   0.0   0.4  0.1   4   20   0  37.4     7 19.50 0.000 2 1.78 4 0019+058 y r g
00 22 32.4420 +06 08 04.290   0.37567832    6.1345249   23.4    11.1    20.6  19.  19.   0.6   1.1  0.1   4   22   0  21.2    23 19.50 0.000 2 1.78 4 0019+058 z r g
00 24 42.9848 -42 02 04.017   0.41194023  -42.0344491   88.2   -55.2   -68.8  74.  74.  -0.7  -0.9  1.3   2   19   0  32.2     5  0.00 0.000 0 9.00 4 0022-423 r r f
\end{verbatim}
\normalsize
\end{center}
\end{table}

\begin{table}
\begin{center}
\caption{Explanation of optical source quality flag (column 24)
         of Tables 4 and 5.}
\vspace*{5mm}
\begin{tabular}{cl}  
flag & description \\
\tableline\tableline
   g & good optical source \\
   f & faint optical source (S/N $\le$ 5) \\
   u & unconfirmed, too faint for unique ID \\
   l & large (opt.-radio) pos.diff. ($\ge  3\sigma$) \\
   p & problem case, ($\ge 5\sigma$) or blended image \\
   n & not an ICRF source \\
\tableline
\end{tabular}
\end{center}
\end{table}

\clearpage

\begin{table}
\begin{center}
\caption{List of empty fields.  These are ICRF sources where no
         optical counterpart could be detected at the radio position.
         Column 1 gives the ICRF source name, column 2 lists the
         observing runs in which the field was observed (see Tab.~1),
         column 3 gives the total number of deep exposures, and the
         last column adds comments, if available.  For some cases
         offset coordinates between an object visible on the deep
         CCD exposures and the ICRF radio position is given in
         arcsec for the RA and Dec component, respectively, in the
         comment column.}
\vspace*{3mm}
\begin{tabular}{clrl}
  source &  runs &  n.exp. &  comment \\
\tableline \tableline
  0008$-$421 & r       &  2 & \\
  0201+088 & z         &  1 & \\
  0334+014 & w         &  6 & $-$21.6 $-$2.5 off bright star \\
  0615$-$365 & t       &  2 & \\
  0637$-$337 & t       &  2 & \\
  0647$-$475 & p,t     &  4 & high uncertainty radio position \\
  0733$-$174 & J       &  5 & \\
  0736$-$332 & s       &  2 & \\
  0826$-$373 & p,s,t,w,o & 31 & offset =  2.3  5.5, calibr.field \\
  0831$-$445 & s,t     & 13 & offset = $-$2.3  0.8,  VELA-G \\
  0833$-$450 & p,s,t   & 16 & offset:   5.0  0.1 \\
  0903$-$573 & J       &  4 & \\
  1039$-$474 & p,s     &  8 & high uncertainty radio position \\
  1148$-$671 & p       &  5 & offset =  0.6  5.0, crowded field \\
  1234$-$504 & J       &  3 & \\
  1236$-$684 & p       &  2 & offset = $-$2.7  3.6 \\
  1251$-$407 & p,s     &  5 & \\
  1334$-$649 & r       &  5 & offset = $-$0.9 $-$3.4 \\
  1352$-$632 & q       &  5 & offset = $-$3.9 $-$4.3, crowded field \\
  1420$-$679 & J       &  7 & \\
  1448$-$648 & J       &  4 & \\
  1508$-$656 & r       &  4 & \\
  1600$-$445 & q       &  8 & offset = $-$1.7  1.3, crowded field \\
  1740$-$517 & q       &  5 & offset =  0.0 $-$8.3 \\
  1822$-$173 & J       &  4 & star within 1 arcsec \\
  1829$-$106 & J       &  6 & \\
  1829$-$718 & q,r     &  5 & offset =  2.5 $-$1.8 \\
  1929+226 & b         & 13 & offset = 1.9  \\
  1932+204 & b         &  2 & \\
  1943+228 & b         &  7 & offset = 3.0, crd field \\
  1950$-$613 & J       &  3 & \\
  2008$-$068 & y,z     &  4 & \\
  2128+048 & y         &  2 & \\
  2259$-$375 & q       &  1 & \\
  2300$-$307 & q,v   &  8 & offset =  1.1 $-$3.8, inside halo of bright star\\
  2333$-$528 & r     &  5 & offset =  2.1  8.2 \\
\tableline
\end{tabular} 
\end{center}
\end{table}

\clearpage

\begin{table}
\begin{center}
\caption{Eulerian orientation angles (w1,w2,w3) between optical and 
         radio coordinate
         system from various least-squares solutions of Table 4 and 5 data.
         Consecutive lines show results from 2 models (with and without
         declination offset term). Results from weighted (w) and unweighted
         (u) solutions as well as with different restrictions (see text) 
         are shown. See text for more explanations.}
\vspace*{5mm}
\begin{verbatim}     
  sigma   ndf    w1    w2    w3 dcoff         errors         notes of
    mas         mas   mas   mas  mas    mas  mas  mas  mas   solution
---------------------------------------------------------------------
  26.75  1240  4.93 -4.78  7.61        1.42 1.34 1.21       u,2, 3sig
  26.55  1239  4.55 -5.11  7.78 -5.14  1.42 1.32 1.20 1.07 
  17.94  1195  6.59 -3.84  6.96        0.98 0.92 0.87       w,2, 3sig
  17.94  1197  6.13 -4.51  6.88 -3.58  0.98 0.93 0.87 0.76 
---------------------------------------------------------------------
  28.98  1261  5.48 -4.71  7.46        1.53 1.44 1.30       u,2, 4sig
  28.58  1259  5.36 -5.06  7.47 -5.31  1.51 1.42 1.28 1.14  
  23.24  1272  3.44 -3.85  7.39        1.20 1.15 1.06       w,2, 4sig
  22.85  1270  3.50 -4.72  7.43 -5.05  1.18 1.14 1.04 0.91
---------------------------------------------------------------------
  29.71  1263  5.10 -3.89  3.92        1.56 1.44 1.37       u,4, 3sig
  29.81  1264  4.60 -3.71  3.92 -3.60  1.56 1.44 1.37 1.19 
  18.44  1184  5.82 -3.33  3.06        1.02 0.95 0.94       w,4, 3sig
  18.45  1185  4.51 -4.22  2.93 -2.84  1.02 0.97 0.94 0.79 
---------------------------------------------------------------------
  33.06  1291  4.15 -4.00  5.50        1.71 1.58 1.50       u,4, 4sig
  32.93  1290  3.83 -4.20  5.50 -4.34  1.71 1.58 1.50 1.30  
  24.63  1287  4.62 -5.70  5.23        1.27 1.19 1.13       w,4, 4sig
  24.65  1287  4.68 -6.19  5.21 -3.15  1.27 1.20 1.13 0.99
---------------------------------------------------------------------
\end{verbatim}
\end{center}
\end{table}

\clearpage

\begin{table}
\begin{center}
\caption{Observations with large (optical$-$radio) position differences
  selected from the UCAC2-type catalog if exceeding 4.0 $\sigma$ in either
  coordinate.  These 88 observations are of 63 distinct sources.
  The format of this table is the same as that of Tables 4 and 5.}
\vspace*{5mm}
\small
\begin{verbatim}
         1           2            3            4         5        6       7    8    9     10    11   12  13   14  15    16    17   18   19  20  21 22    23  24 25 26  27
    RAJ2000      DEJ2000        RAJ2000      DEJ2000S   dRAc    dDec    dist  eRA  eDE   sRA   sDE eRad Nccd  Nr Nrx   err   S/N  mag    z   T   S  U   name Run F Q  notes
 h  m  s        d  '   "         degree       degree     mas     mas     mas  mas  mas              mas                mas
00 10 31.0016 +10 58 29.484   0.17527823   10.9748568   66.0   -62.9   -19.9  13.  13.  -4.8  -1.5  0.1   4   15   3  28.1    11 15.40 0.089 1 1.82 2 0007+106 z r l  galaxy                     
01 50 02.6937 -07 25 48.349   1.83408158   -7.4300970  148.5   -53.2   138.6  17.  17.  -3.1   8.0  3.5   3   21   0  19.0    47 15.62 0.017 1 9.00 2 0147-076 v r p  compact galaxy             
01 50 02.6931 -07 25 48.342   1.83408143   -7.4300949  158.4   -61.3   146.1  20.  20.  -3.0   7.2  3.5   3   19   0  25.0   156 15.62 0.017 1 9.00 2 0147-076 z r p  compact galaxy             
02 10 46.1985 -51 01 01.846   2.17949959  -51.0171794   49.4   -17.9    46.0  10.  10.  -1.8   4.6  0.7   5   28   2  18.3    45 16.93 1.003 2 3.59 2 0208-512 r r l                             
02 17 48.9506 +01 44 49.672   2.29693071    1.7471312   68.4   -62.9   -26.8   9.   9.  -7.0  -3.0  0.1   5   22   0  22.4   757 16.09 1.715 3 1.72 2 0215+015 w r p                             
02 17 48.9508 +01 44 49.692   2.29693077    1.7471366   60.1   -59.7    -7.3  13.  13.  -4.6  -0.6  0.1   4   23   2  24.2    27 16.09 1.715 3 1.72 2 0215+015 z r l                             
02 40 08.1788 -23 09 15.758   2.66893855  -23.1543772   65.9    59.5   -28.3   9.   9.   6.6  -3.1  0.1   3   16   0  12.7   110 16.63 2.225 3 5.75 2 0237-233 v r p                             
02 41 04.8027 -08 15 20.840   2.68466741   -8.2557889  107.8    62.0   -88.2  19.  19.   3.3  -4.6  0.8   4   26   0  19.7   260 12.31 0.005 1 4.36 2 0238-084 v r l  extended galaxy            
02 41 04.8010 -08 15 20.865   2.68466694   -8.2557958  118.9    36.8  -113.1  17.  17.   2.2  -6.6  0.8   4   26   1  29.9   600 12.31 0.005 1 4.36 2 0238-084 z r p  galaxy                     
02 53 29.1745 -54 41 51.472   2.89143735  -54.6976312   63.1   -51.5   -36.4  10.  10.  -5.2  -3.6  0.2   3   32   3  18.3   159 17.80 0.537 3 9.00 2 0252-549 r r p                             
03 03 50.6211 -62 11 25.579   3.06406141  -62.1904385   77.4   -71.9   -28.7  11.  11.  -6.5  -2.6  0.1   4   26   0  21.1    59 17.87 1.351 3 9.00 2 0302-623 o r p                             
03 09 56.0919 -60 58 39.073   3.16558108  -60.9775204   55.7   -52.9   -17.3   9.   9.  -5.9  -1.9  0.1   6   27   1  13.9    28 18.18 1.480 3 1.40 2 0308-611 o r p                             
03 40 35.6051 -21 19 31.193   3.67655698  -21.3253313   43.4   -38.2   -20.7   8.   8.  -4.8  -2.6  0.1   6   26   1  20.3   107 17.10 0.223 2 3.73 2 0338-214 w r l                             
04 07 48.4303 -12 11 36.702   4.13011953  -12.1935282   43.3    -9.8   -42.2   7.   7.  -1.4  -6.0  0.1   6   32   0  16.1   686 14.86 0.574 3 3.14 2 0405-123 w r p                             
04 09 05.7690 -12 38 48.186   4.15160249  -12.6467183   43.5   -11.1   -42.1  10.  10.  -1.1  -4.2  0.1   5   39   0  16.1    37 18.60 1.563 3 3.13 2 0406-127 w r l                             
04 57 03.1771 -23 24 52.057   4.95088254  -23.4144602   46.5   -28.7   -36.6   9.   9.  -3.2  -4.1  0.0   5   38   1  16.4    64 18.90 1.003 3 2.21 2 0454-234 w r l                             
05 22 57.9819 -36 27 30.870   5.38277275  -36.4585749   38.0   -33.1   -18.6   8.   8.  -4.1  -2.3  0.1   5   48   1  14.5   903 14.62 0.055 1 3.32 2 0521-365 o r l                             
05 25 31.4050 -45 57 54.708   5.42539029  -45.9651966   55.9    51.0   -22.9  11.  11.   4.6  -2.1  0.2   2   53   2  18.9    98 18.00 1.479 3 9.00 2 0524-460 p r l                             
05 25 31.4040 -45 57 54.681   5.42539001  -45.9651892   40.7    40.5     3.7  10.  10.   4.1   0.4  0.2   4   52   0  25.1    89 18.00 1.479 3 9.00 2 0524-460 s r l                             
05 38 50.3622 -44 05 08.970   5.64732284  -44.0858249   31.5     7.2   -30.7   7.   7.   1.0  -4.4  0.0   6   45   2  18.4   153 21.00 0.885 1 2.63 2 0537-441 o r l                             
06 16 35.9852 -34 56 16.704   6.27666255  -34.9379732  155.8    56.8  -145.1  33.  33.   1.6  -4.2 10.5   3   80   4  22.7    11  0.00 0.000 0 9.00 2 0614-349 w r l                             
06 35 46.5109 -75 16 16.781   6.59625304  -75.2713281   36.1    11.6    34.2   6.   6.   1.9   5.7  0.5   5   71   3  18.8   202 15.75 0.651 3 4.30 2 0637-752 o r l                             
06 48 14.0941 -30 44 19.670   6.80391504  -30.7387971   31.6   -30.0    -9.9   5.   5.  -6.0  -2.0  0.1  24  223  15  15.8    30 20.40 1.153 3 3.23 2 0646-306 o c p                             
06 50 24.6075 -16 37 39.995   6.84016876  -16.6277763  457.0   369.1  -269.5  30.  30.  12.3  -9.0  0.1   3  171   7  22.3    10  0.00 0.000 0 1.91 2 0648-165 w r p                             
06 50 24.6101 -16 37 40.035   6.84016947  -16.6277876  510.8   405.8  -310.2  70.  70.   5.8  -4.4  0.1   2  202   4  34.6     7  0.00 0.000 0 1.91 2 0648-165 x r p                             
07 29 05.3214 -36 39 46.687   7.48481150  -36.6629687 1810.1 -1093.2 -1442.7  15.  15. -67.8 -89.5  5.9   4  246  13  21.5     2  0.00 0.000 0 3.66 2 0727-365 p r p  **6px flr=0.1              
07 29 05.3197 -36 39 46.685   7.48481103  -36.6629680 1820.5 -1113.6 -1440.2  16.  16. -65.3 -84.4  5.9   4  222  15  22.2     2  0.00 0.000 0 3.66 2 0727-365 t r p  foreground star            
08 20 57.4444 -12 58 59.156   8.34929012  -12.9830988   48.4   -46.5    13.4  11.  11.  -4.2   1.2  0.1   5  194   7  28.2    77 15.00 0.000 2 3.47 2 0818-128 w r l                             
08 20 57.4450 -12 58 59.167   8.34929028  -12.9831020   38.1   -38.1     1.9   8.   8.  -4.8   0.2  0.1   4  187   7  23.5   178 15.00 0.000 2 3.47 2 0818-128 x r l                             
10 35 02.1520 -20 11 34.342  10.58393111  -20.1928729   49.5   -46.5    17.1  11.  11.  -4.2   1.6  0.1   7   46   3  20.6    22 18.20 2.198 3 3.19 2 1032-199 t r l                             
11 30 07.0487 -14 49 27.349  11.50195797  -14.8242637   68.6   -56.6    38.8   9.   9.  -6.3   4.3  0.1   5   36   0  24.2    70 16.90 1.187 3 4.27 2 1127-145 t r p                             
11 30 07.0504 -14 49 27.376  11.50195845  -14.8242711   33.8   -31.5    12.2   7.   7.  -4.5   1.7  0.1   5   37   1  18.3   142 16.90 1.187 3 4.27 2 1127-145 u r l                             
11 33 20.0318 +00 40 52.880  11.55556440    0.6813556  361.9  -359.3    42.9  49.  49.  -7.3   0.9  0.1   6   24   0  27.5     5 19.43 1.633 3 2.61 2 1130+009 x r p                             
12 18 06.2482 -46 00 28.763  12.30173562  -46.0079896  251.3   -41.9   247.8  20.  20.  -1.9  11.0 10.1   5  166   6  18.2    12 20.30 0.529 1 9.00 2 1215-457 s r p                             
12 18 06.2505 -46 00 28.761  12.30173626  -46.0079893  249.5   -17.9   248.9  33.  33.  -0.5   7.2 10.1   4  162   6  19.3     7 20.30 0.529 1 9.00 2 1215-457 t r p                             
12 20 11.8869 +02 03 42.176  12.33663526    2.0617155   60.7    35.2   -49.4  10.  10.   3.5  -4.9  0.3   3   19   0  30.8   326 15.97 0.240 3 9.00 2 1217+023 x r l                             
12 25 03.7405 +12 53 13.029  12.41770569   12.8869526  117.5   -41.7  -109.9  22.  22.  -1.9  -5.0  0.2   3   16   0  21.7   373 12.31 0.003 1 2.16 2 1222+131 y r p  bright galaxy              
12 56 11.1688 -05 47 21.533  12.93643577   -5.7893148   34.0    32.9    -8.6   8.   8.   4.1  -1.1  1.3   6   36   0  22.5   370 17.75 0.538 3 4.05 2 1253-055 y r l                             
13 05 33.0125 -10 33 19.424  13.09250347  -10.5553956   37.7   -37.5     3.9   8.   8.  -4.7   0.5  0.1   6   28   0  24.5   310 15.23 0.286 3 3.30 2 1302-102 u r l                             
13 47 33.4536 +12 17 23.887  13.79262601   12.2899686 1394.2  1348.7  -353.3  53.  53.  25.4  -6.7  0.3   3   12   0  27.6    18 18.44 0.120 1 5.35 2 1345+125 x r p  galaxy                     
14 04 45.8918 -01 30 21.927  14.07941440   -1.5060908   58.1   -54.5    20.1  12.  12.  -4.5   1.7  0.1   4   26   0  26.2    35 18.45 2.522 3 2.38 2 1402-012 x r l                             
14 32 57.6945 -18 01 35.252  14.54935959  -18.0264588   55.8    55.7    -3.1  12.  12.   4.6  -0.3  0.2   5   61   2  24.2    37 18.70 2.331 3 3.87 2 1430-178 u r l                             
14 38 09.4643 -22 04 54.727  14.63596231  -22.0818686   73.8   -70.6    21.5  12.  12.  -5.9   1.8  0.1   4   74   4  24.4    68 18.10 1.187 3 4.50 2 1435-218 u r p                             
14 54 27.4064 -37 47 33.154  14.90761290  -37.7925429   40.5   -39.3    -9.6   8.   8.  -4.9  -1.2  0.1   6  209   8  19.5    52 16.69 0.314 3 2.93 2 1451-375 t c l                             
15 05 06.4757 +03 26 30.757  15.08513215    3.4418770   59.4   -21.2   -55.5  12.  12.  -1.8  -4.6  0.1   4   38   1  28.6    36 18.62 0.409 3 2.78 2 1502+036 x r l                             
15 17 41.8145 -24 22 19.439  15.29494847  -24.3720664   41.4    18.6    37.0   8.   8.   2.3   4.6  0.1   5   78   2  20.8   460 14.80 0.048 2 3.67 2 1514-241 u r l                             
15 34 52.4594 +01 31 04.080  15.58123872    1.5178000  152.8    85.6  -126.6  25.  25.   3.4  -5.1  0.1   4   48   4  31.9    10 19.69 1.420 3 4.13 2 1532+016 y r l                             
15 49 29.4374 +02 37 01.114  15.82484372    2.6169761   50.2     8.2   -49.5   9.   9.   0.9  -5.5  0.1   8   39   2  34.6    37 17.42 0.412 3 2.92 2 1546+027 b r l                             
15 49 29.4369 +02 37 01.119  15.82484358    2.6169776   44.1     0.7   -44.1   8.   8.   0.1  -5.5  0.1   4   46   2  27.8   102 17.42 0.412 3 2.92 2 1546+027 x r p                             
15 49 29.4371 +02 37 01.113  15.82484365    2.6169758   50.7     4.5   -50.5  11.  11.   0.4  -4.6  0.1   5   55   2  27.5    51 17.42 0.412 3 2.92 2 1546+027 y r l                             
15 50 35.2739 +05 27 10.441  15.84313163    5.4529004   69.5    69.1    -7.0  13.  13.   5.3  -0.5  0.1   8   20   0  24.2    16 18.73 1.422 3 2.91 2 1548+056 b r l                             
15 50 35.2736 +05 27 10.437  15.84313155    5.4528991   65.8    64.8   -11.7  15.  15.   4.3  -0.8  0.1   3   39   1  24.8    26 18.73 1.422 3 2.91 2 1548+056 x r l                             
16 16 37.5511 +04 59 32.744  16.27709752    4.9924290   86.1   -85.8     7.7  21.  21.  -4.1   0.4  0.1   3   51   3  24.1    13 19.17 3.217 3 3.11 2 1614+051 x r l                             
17 23 41.0173 -65 00 36.676  17.39472702  -65.0101878  100.5   -76.8   -64.8  16.  16.  -4.8  -4.0  0.2  10  297  20  23.6   100 15.50 0.014 1 5.40 2 1718-649 q r l  galaxy structure           
17 33 02.7345 -13 04 49.502  17.55075959  -13.0804171  422.5   419.9    46.6  38.  38.  11.1   1.2  0.1   4  213  10  21.6    12 19.50 0.902 3 2.73 2 1730-130 y r p  **5px flr=2                
17 37 35.7561 -56 34 03.278  17.62659893  -56.5675771  167.5  -116.6  -120.2  18.  18.  -6.4  -6.6  3.2   5  375  29  19.4    45 17.00 0.098 1 9.00 2 1733-565 q r p  faint galaxy               
18 19 35.2196 -63 45 48.551  18.32644988  -63.7634864 1479.4  1438.5  -345.3  14.  14. 102.3 -24.6  1.3   2  190   8  21.4   207 16.00 0.065 1 5.50 2 1814-637 q r p  elongated galaxy           
18 24 02.8561 +10 44 23.807  18.40079335   10.7399465   35.5    11.9    33.4   7.   7.   1.7   4.8  0.1  12  168  10  24.7    33 17.27 1.364 3 3.08 2 1821+107 b r l  1 B filter                 
19 11 09.6579 -20 06 55.148  19.18601607  -20.1153188   79.9    69.9   -38.7   9.   9.   7.8  -4.3  0.0   5  421  22  23.9    75 18.10 1.119 3 2.36 2 1908-201 u r p                             
19 11 09.6565 -20 06 55.136  19.18601570  -20.1153155   57.7    51.1   -26.8   9.   9.   5.7  -3.0  0.0   5  351  21  25.1   110 18.10 1.119 3 2.36 2 1908-201 y r p                             
19 24 51.0585 -29 14 30.143  19.41418291  -29.2417064   39.7    33.0   -22.0   8.   8.   4.1  -2.7  0.0   6  275  11  27.5   116 18.21 0.352 3 2.61 2 1921-293 u c l                             
19 28 40.8376 +08 48 47.505  19.47801044    8.8131958  946.1  -265.6  -908.1 200. 200.  -1.3  -4.5  0.2   1  261  21  22.9     2  0.00 0.000 0 2.97 2 1926+087 z r u                             
19 39 24.9686 -63 42 45.674  19.65693572  -63.7126872  386.3  -383.1   -49.4  63.  63.  -6.1  -0.8  2.0   4   78   5  18.4     4 18.40 0.183 1 6.40 2 1934-638 r r p                             
19 39 57.2567 -10 02 41.546  19.66590463  -10.0448739   25.4     1.4   -25.4   4.   4.   0.3  -6.4  0.1  13  237  11  20.0   168 17.00 3.787 3 3.40 2 1937-101 y c p                             
20 03 24.1136 -32 51 45.189  20.05669822  -32.8625524   65.7   -34.6   -55.8   8.   8.  -4.3  -7.0  0.1   5  109   5  22.8    75 17.30 3.783 3 4.08 2 2000-330 q r p                             
20 03 24.1136 -32 51 45.191  20.05669821  -32.8625531   68.1   -35.1   -58.3   8.   8.  -4.4  -7.3  0.1   5  109   4  27.5   180 17.30 3.783 3 4.08 2 2000-330 u r p                             
20 11 15.7076 -15 46 40.301  20.18769656  -15.7778613   67.1   -47.9   -47.0   9.   9.  -5.3  -5.2  0.1   4  116   3  22.1    60 18.30 1.180 3 1.95 2 2008-159 u r p                             
20 11 15.7073 -15 46 40.302  20.18769648  -15.7778616   70.8   -52.0   -48.1   9.   9.  -5.8  -5.3  0.1   5   99   2  20.9   159 18.30 1.180 3 1.95 2 2008-159 y r p                             
21 01 38.8313 +03 41 31.309  21.02745315    3.6920304   43.8   -42.3   -11.5   7.   7.  -6.0  -1.6  0.1  10   91   2  18.9    65 17.78 1.015 3 2.70 2 2059+034 y c p                             
21 01 38.8311 +03 41 31.322  21.02745308    3.6920340   46.0   -46.0     1.4   7.   7.  -6.6   0.2  0.1  13   88   3  20.5    35 17.78 1.015 3 2.70 2 2059+034 z c p                             
21 16 30.2496 -80 53 54.316  21.27506933  -80.8984210 1680.9 -1414.9   907.5 100. 100. -14.1   9.1  0.2   3   58   2  20.7     7  0.00 0.000 0 3.60 2 2109-811 q r p  faint comp. in triple      
21 31 35.2591 -12 07 04.792  21.52646085  -12.1179979   39.7   -39.5     3.6   9.   9.  -4.4   0.4  1.1   5   54   1  20.1   260 16.11 0.501 3 4.32 2 2128-123 u r l                             
21 31 35.2591 -12 07 04.788  21.52646086  -12.1179966   39.8   -38.9     8.3   7.   7.  -5.5   1.2  1.1   4   52   1  17.2   191 16.11 0.501 3 4.32 2 2128-123 v r p                             
21 31 35.2583 -12 07 04.800  21.52646063  -12.1180001   51.3   -51.1    -4.3   5.   5. -10.0  -0.8  1.1   8   47   1  20.4   162 16.11 0.501 3 4.32 2 2128-123 z r p                             
22 18 52.0362 -03 35 36.830  22.31445449   -3.5935640   54.4   -23.4    49.1   8.   8.  -2.9   6.1  0.1   3   34   0  19.9   108 16.38 0.901 3 2.89 2 2216-038 v r p                             
22 18 52.0363 -03 35 36.836  22.31445453   -3.5935655   48.6   -21.2    43.7   7.   7.  -3.0   6.2  0.1   5   29   0  24.8    99 16.38 0.901 3 2.89 2 2216-038 y r p                             
22 18 52.0341 -03 35 36.835  22.31445393   -3.5935654   69.3   -53.5    44.0  12.  12.  -4.5   3.7  0.1   4   34   0  23.6    46 16.38 0.901 3 2.89 2 2216-038 z r l                             
22 25 47.2540 -04 57 01.354  22.42979279   -4.9503760   86.8   -78.4    37.2  17.  17.  -4.6   2.2  0.0   3   24   2  26.1    28 18.39 1.404 3 2.06 2 2223-052 u r l                             
22 25 47.2551 -04 57 01.381  22.42979307   -4.9503836   64.2   -63.4     9.8  10.  10.  -6.3   1.0  0.0   4   28   2  21.1    58 18.39 1.404 3 2.06 2 2223-052 v r p                             
22 25 47.2553 -04 57 01.385  22.42979315   -4.9503846   59.4   -59.1     6.2  10.  10.  -5.9   0.6  0.0   2   33   0  23.4   118 18.39 1.404 3 2.06 2 2223-052 z r p                             
22 29 40.0842 -08 32 54.511  22.49446782   -8.5484753   75.8    -2.8   -75.7  13.  13.  -0.2  -5.8  0.1   2   24   1  16.7    45 17.41 1.562 3 1.95 2 2227-088 u r p                             
22 29 40.0836 -08 32 54.507  22.49446766   -8.5484743   73.0   -11.3   -72.1  10.  10.  -1.1  -7.2  0.1   4   26   0  16.6    54 17.41 1.562 3 1.95 2 2227-088 v r p                             
22 29 40.0826 -08 32 54.504  22.49446740   -8.5484732   72.6   -25.2   -68.1  14.  14.  -1.8  -4.9  0.1   4   26   1  21.0    36 17.41 1.562 3 1.95 2 2227-088 z r l                             
22 58 05.9604 -27 58 21.276  22.96832234  -27.9725768   38.1   -32.6   -19.7   7.   7.  -4.7  -2.8  0.0   5   23   1  23.8   360 16.77 0.927 3 2.01 2 2255-282 u r l                             
23 14 48.4955 -31 38 39.598  23.24680430  -31.6443328   96.8   -65.2   -71.6  12.  12.  -5.4  -6.0  0.1   5   19   0  28.0    61 18.30 1.322 3 3.10 2 2312-319 r r p                             
23 14 48.4958 -31 38 39.558  23.24680439  -31.6443216   68.6   -61.0   -31.3   8.   8.  -7.6  -3.9  0.1   6   18   1  14.0    70 18.30 1.322 3 3.10 2 2312-319 u r p                             
23 31 59.4822 -38 11 47.724  23.53318950  -38.1965901  102.8    71.5   -73.9  14.  14.   5.1  -5.3  0.1   3   17   0  21.4    29 17.04 1.195 3 1.95 2 2329-384 q r p                             
23 37 57.3396 -02 30 57.702  23.63259434   -2.5160283   73.2     8.3   -72.7  17.  17.   0.5  -4.3  0.1   3    9   0  22.8    26 19.60 1.072 3 3.29 2 2335-027 y r l                             
\end{verbatim}
\normalsize
\end{center}
\end{table}

\clearpage
\newpage

\begin{table}
\begin{center}
\caption{List of nearest neighbor sources in our sample.}
\vspace*{5mm}
\begin{tabular}{cccrrrrrrc}
separation & \multicolumn{2}{c}{ICRF name}
   & \multicolumn{4}{c}{optical$-$radio source 1, 2}
   & \multicolumn{2}{c}{pos.error} & Fig.\\
   & source1 & source2 &  RA  & Dec & RA  & Dec & src.1 & src.2 & numb.\\
  (degree)  &    &    & (mas)&(mas)&(mas)&(mas)& (mas) & (mas) &      \\ 
\tableline\tableline
 0.556 & 0405$-$123 & 0406$-$127 &  -9 & -45 &  -5 & -38 & 12 & 12 &10\\
 0.569 & 1042+071   & 1043+066   &  -6 & -32 & -32 &  -9 & 78 & 10 &  \\
 0.608 & 1443$-$162 & 1445$-$161 &  -6 & +24 & +22 & +42 & 17 & 15 &11\\
 0.752 & 0503$-$608 & 0506$-$612 &  +6 & +18 &  +9 & -12 & 59 & 10 &  \\
 0.943 & 1219+044   & 1222+037   & -18 & -10 &  +7 & +26 & 10 & 28 &  \\
 0.977 & 0457+024   & 0500+019   &  +2 & -28 & +92 &-313 & 15 &250 &  \\
 0.992 & 0422$-$380 & 0426$-$380 & -20 & -25 & -18 & +11 & 11 & 17 &  \\
 1.009 & 0403$-$132 & 0406$-$127 &  +6 & +20 &  -5 & -38 & 12 & 12 &10\\
 1.123 & 1510$-$089 & 1511$-$100 &  -2 &  +4 & -11 & +11 & 12 & 11 &  \\
 1.157 & 0738$-$674 & 0743$-$673 &  +2 &  -8 &  +7 &  -5 & 12 &  6 &  \\
 1.173 & 1038+064   & 1043+066   & -10 & -30 & -32 &  -9 & 12 & 10 &12\\
\tableline
\end{tabular}
\end{center}
\end{table}

\end{document}